\documentclass[article,10pt]{IEEEtran}

\usepackage{cite}

\usepackage{amsmath,amssymb,amsthm,bm}
\usepackage{mathtools}

\usepackage{algorithm2e}
\usepackage{microtype}

\usepackage{algorithmic}

\usepackage{array}

\usepackage{url}

\usepackage[utf8]{inputenc}
\usepackage[T1]{fontenc}
\usepackage{pgf}
\usepackage{xcolor}
\usepackage{physics}
\usepackage{notation}
\usepackage{multirow}
\usepackage{amssymb}
\usepackage{acronym}
\usepackage{psfrag}
\usepackage{soul}
\usepackage{import}
\usepackage[inline]{enumitem}
\usepackage{tikz}
\usepackage{pgfplots}

\usetikzlibrary{decorations.text}
\usetikzlibrary{decorations.shapes}
\usepgfplotslibrary{patchplots}
\usepgfplotslibrary{colormaps}
\usepgfplotslibrary{groupplots}
\usetikzlibrary{calc}
\usetikzlibrary{decorations.pathreplacing,decorations.markings,shapes.geometric}
\usetikzlibrary{decorations.pathmorphing}
\usetikzlibrary{fit}

\usetikzlibrary{calc,arrows.meta}

\usetikzlibrary{backgrounds} % scopes is needed for on background layer syntax

\newlength{\figureheight}
\newlength{\figurewidth}
\graphicspath{{./figures/}}

\newcommand{\exportFigures}{true}
\newcommand{\exportFiguresAsPNG}{true}

\ifthenelse{\equal{\exportFigures}{true}}
{
	\usepgfplotslibrary{external}
	\tikzexternalize[prefix=compiled_tikz_figures/,optimize command away=\includepdf]
	\ifthenelse{\equal{\exportFiguresAsPNG}{true}}
	{
		\tikzset
		{   png export/.style={
				external/system call={
					pdflatex \tikzexternalcheckshellescape -halt-on-error --extra-mem-top=10000000 -interaction=batchmode -jobname "\image" "\texsource" && pdftops -eps "\image.pdf" && convert -density 700 -transparent white "\image.pdf" "\image.png"
		}}}
		\tikzset{png export}
	}
	{}
}
{}

\definecolor{mycolor01}{rgb}{0.00000,0.00000,1.00000}
\definecolor{mycolor02}{rgb}{0.133,0.545,0.133}
\definecolor{mycolor03}{rgb}{0.50000,0.00000,0.50000}
\definecolor{mycolor05}{rgb}{1.00000,0.83984,0.00000}
\definecolor{mycolor04}{rgb}{0.92969,0.50781,0.92969}
\definecolor{mycolor06}{rgb}{1.00000,0.64453,0.00000}
\definecolor{mycolor07}{rgb}{0.50000,0.50000,0.50000}
\definecolor{mycolor08}{rgb}{1.00000,0.00000,0.00000}
\definecolor{mycolor09}{rgb}{0.2510 ,0.8784, 0.8157}
\definecolor{mycolor10}{rgb}{0.54297,0.00000,0.00000}
\definecolor{mycolor11}{rgb}{0.6445, 0.1641,0.1641}
\definecolor{mycolor12}{rgb}{1, 0, 1}

\pgfdeclarelayer{back1}
\pgfdeclarelayer{lay1}
\pgfdeclarelayer{back2}
\pgfdeclarelayer{lay2}
\pgfsetlayers{back2,lay2,back1,main,lay1}

\makeatletter

\tikzset{
	nomorepostactions/.code={\let\tikz@postactions=\pgfutil@empty},
	decmark/.style 2 args={decoration={markings,
			mark= between positions 0 and 1 step (1/6)*\pgfdecoratedpathlength with{%
				\tikzset{#2,every mark}\tikz@options
				\pgftransformresetnontranslations
				\pgfuseplotmark{#1}%
			},  
		},
		postaction={decorate},
		/pgfplots/legend image post style={
			mark=#1, mark options={#2}, every path/.append style={nomorepostactions}
		},
	},
	markbeginend/.style 2 args={decoration={markings,
			mark= between positions 0 and 1 step (1)*\pgfdecoratedpathlength with{%
				\tikzset{#2,every mark}\tikz@options
				\pgfuseplotmark{#1}%
			},  
		},
		postaction={decorate},
		/pgfplots/legend image post style={
			mark=#1,mark options={#2},every path/.append style={nomorepostactions}
		},
	},
	markend/.style 2 args={decoration={markings,
			mark= at position \pgfdecoratedpathlength with{%
				\tikzset{#2,every mark}\tikz@options
				\pgfuseplotmark{#1}%
			},  
		},
		postaction={decorate},
		/pgfplots/legend image post style={
			mark=#1,mark options={#2},every path/.append style={nomorepostactions}
		},
	},
	posmark/.style 2 args={decoration={markings,
			mark= at position #2 with{%
				\tikzset{solid,every mark}\tikz@options
				\pgftransformresetnontranslations
				\pgfuseplotmark{#1}%
			},  
		},
		postaction={decorate},
		/pgfplots/legend image post style={
			mark=#1,mark options={solid},every path/.append style={nomorepostactions}
		},
	},
}

\makeatother

\pgfplotsset{
resultStyle1/.style={mark=none, line width=0.5pt, mycolor01, decmark={oplus}{solid}},
resultStyle2/.style={mark=none, line width=0.5pt, mycolor02, decmark={triangle}{solid}},%{draw=black,fill=yellow,solid}},
resultStyle3/.style={mark=none ,line width=0.5pt, mycolor03, decmark={+}{solid}},
resultStyle4/.style={mark=none, line width=0.5pt, mycolor06, decmark={star}{solid}},
resultStyle5/.style={mark=none, line width=0.5pt, mycolor08, decmark={o}{solid}},
resultStyle6/.style={mark=none, line width=0.5pt, mycolor05, decmark={square}{solid}}, 
resultStyle7/.style={mark=none, line width=0.5pt, mycolor09, decmark={diamond}{solid}}, 
resultStyle8/.style={mark=none, line width=0.5pt, mycolor11, decmark={otimes}{solid}}, 
resultStyle9/.style={mark=none, line width=0.5pt, mycolor12, decmark={x}{solid}}, 
resultStyleBase/.style={mark=none, line width=0.5pt,}, 
compareStyle1/.style={mark=none, line width=0.5pt, mycolor01},
compareStyle2/.style={mark=none, line width=0.5pt, mycolor02},%{draw=black,fill=yellow,solid}},
compareStyle3/.style={mark=none ,line width=0.5pt, mycolor03},
compareStyle4/.style={mark=none, line width=0.5pt, mycolor06},
compareStyle5/.style={mark=none, line width=0.5pt, mycolor08},
compareStyle6/.style={mark=none, line width=0.5pt, mycolor05}, 
compareStyle7/.style={mark=none, line width=0.5pt, mycolor09}, 
compareStyle8/.style={mark=none, line width=0.5pt, mycolor11}, 
compareStyle9/.style={mark=none, line width=0.5pt, mycolor12}, 
}

\pgfplotsset{
compat=newest,
simple style group/.style={
label style={font=\scriptsize},
legend style={font=\scriptsize},
tick label style={font=\scriptsize},
nodes near coords style={font=\scriptsize},
title style={font=\scriptsize},
scale only axis,
grid style={dotted},
mark options={solid}, % needs to be at the end
},
simple style/.style={
label style={font=\scriptsize},
legend style={font=\scriptsize},
tick label style={font=\scriptsize},
nodes near coords style={font=\scriptsize},
title style={font=\scriptsize},
width=\figurewidth,
height=\figureheight,
at={(0\figurewidth,0\figureheight)},
scale only axis,
grid style={dotted},
mark options={solid}, % needs to be at the end
},
base style/.style={
label style={font=\scriptsize},
legend style={font=\scriptsize},
tick label style={font=\scriptsize},
nodes near coords style={font=\scriptsize},
title style={font=\scriptsize},
width=\figurewidth,
height=\figureheight,
at={(0\figurewidth,0\figureheight)},
scale only axis,
cycle list={
	{mark=none, line width=0.5pt, mycolor01, solid},
	{mark=none, line width=0.5pt, mycolor02, dash dot},
	{mark=none ,line width=0.5pt, mycolor03, densely dashed},
	{mark=none, line width=0.5pt, mycolor04, dash dot dot},
	{mark=x   , line width=0.5pt, mycolor05},
	{mark=.   , line width=0.7pt, mycolor06}, 
	{mark=square,only marks, mark size = 0.8pt, mycolor07,
		mark options = {line width = 0.4pt}},
	{mark=x,     only marks, mark size = 1.3pt, mycolor08,
		mark options = {line width = 0.4pt}},
	{mark=o,     only marks, mark size = 0.8pt, mycolor09,
		mark options = {line width = 0.4pt}},
	{mark=o, mycolor10},
},
grid style={dotted},
xmajorgrids,
ymajorgrids,
mark options={solid}, % needs to be at the end
},
base style group/.style={
label style={font=\scriptsize},
legend style={font=\scriptsize},
tick label style={font=\scriptsize},
nodes near coords style={font=\scriptsize},
title style={font=\scriptsize},
scale only axis,
grid style={dotted},
xmajorgrids,
ymajorgrids,
mark options={solid}, % needs to be at the end
},
std graph style new/.style={
xlabel style={yshift=1mm},
ylabel style={yshift=-1.5mm},
yticklabel style={xshift=1mm},
},
color lines style/.style={
cycle list={
	{mark=none, mycolor01, decmark={oplus}{solid} },
	{mark=none, mycolor02, decmark={+}{solid} }, 
	{mark=none, mycolor03, decmark={triangle}{solid} }, 
	{mark=none, mycolor04, decmark={star}{solid} }, 
	{mark=none, mycolor05, decmark={o}{solid} },
	{mark=none, mycolor06, decmark={square}{solid} },
},
},
meas graph style/.style={
xlabel style={yshift=1mm},
ylabel style={yshift=-1mm},
xmajorgrids,
ymajorgrids,
mark repeat = 1,
mark phase = 0,
cycle list={
	{color=black, only marks, mark=*, mark size=0.5pt, mark options={solid, black}},
	{color=red, only marks, mark=*, mark size=0.1pt, line width=0.25pt},
},
ylabel={},
}, 
ci graph style/.style={
xlabel style={yshift=1mm},
ylabel style={yshift=-1.5mm},
yticklabel style={xshift=1mm},
mark repeat = 1,
mark phase = 0,
ymin=1e-3,
ymax=100,
ytick = {100, 50, 10, 1, 0.1, 0.01, 1e-3, 1e-4},
yticklabels = {$0$, $50$, $90$, $99$, $99.9$, $99.99$, $99.999$, $99.9999$},
y dir=reverse,
},     
bp coeff style/.style={
scale only axis=true,
width=0.225*.9\linewidth,
height=0.225*.9\linewidth,
scale only axis,
xmin=-4.000,
xmax=4.000,
xlabel={$\ell${\color{white}$\aod$}},
ticklabel style={font=\footnotesize},
ymin=0.000, ymax=0.9,
ylabel={$c_\ell$},
xlabel style={font=\footnotesize},
ylabel style={font=\footnotesize},
major tick length=2pt%,
},
}

\acrodef{rf}[RF]{radio-frequency}
\acrodef{mt}[MT]{mobile terminal}
\acrodef{mimo}[MIMO]{multiple-input multiple-output}
\acrodef{pmf}[PMF]{probability mass function}
\acrodef{pdf}[PDF]{probability density function}
\acrodef{cdf}[CDF]{cumulative distribution function}
\acrodef{mmse}[MMSE]{minimum mean-square error}
\acrodef{da}[DA]{data association}
\acrodef{bp}[BP]{belief propagation}
\acrodef{spa}[SPA]{sum-product algorithm}
\acrodef{fg}[FG]{Factor Graph}
\acrodef{nebp}[NEBP]{neural enhanced belief propagation}
\acrodef{mot}[MOT]{multi-object tracking}
\acrodef{jpda}[JPDA]{joint probabilistic data association}
\acrodef{mht}[MHT]{multiple hypothesis tracker}
\acrodef{rfs}[RFS]{random finite sets}
\acrodef{gospa}[GOSPA]{generalized optimal sub-pattern assignment}
\acrodef{rmse}[RMSE]{root mean square error}
\acrodef{mpc}[MPC]{multipath component}
\acrodef{dmc}[DMC]{dense multipath component}
\acrodef{slam}[SLAM]{simultaneous localization and mapping}
\acrodef{mpslam}[MP-SLAM]{multipath-based simultaneous localization and mapping}
\acrodef{2d}[2-D]{two dimensional}
\acrodef{pa}[PA]{physical anchor}
\acrodef{bs}[BS]{base station}
\acrodef{va}[VA]{virtual anchor}
\acrodef{los}[LOS]{line-of-sight}
\acrodef{nlos}[NLOS]{non-line-of-sight}
\acrodef{olos}[OLOS]{obstructed-line-of-sight}
\acrodef{iid}[i.i.d.]{independent and identically distributed}
\acrodef{2d}[2-D]{two-dimensional}
\acrodef{snr}[SNR]{signal-to-noise ratio}
\acrodef{tbd}[TBD]{track-before-detect}
\acrodef{crlb}[CRLB]{Cram\'er-Rao lower bound}
\acrodef{toa}[ToA]{time-of-arrival}
\acrodef{doa}[DoA]{direction-of-arrival}
\acrodef{dod}[DoD]{direction-of-departure}
\acrodef{nn}[NN]{neural network}
\acrodef{dnn}[DNN]{deep neural network}
\acrodef{ne}[NE]{neural-enhanced}
\acrodef{ai}[AI]{artificial intelligence}
\acrodef{mlp}[MLP]{multi-layer perceptron}
\acrodef{ml}[ML]{machine learning}
\acrodef{em}[EM]{expectation maximization}
\acrodef{elbo}[ELBO]{evidence lower bound}
\acrodef{kld}[KLD]{Kullback--Leibler divergence}
\acrodef{imu}[IMU]{inertial measurement unit}

\newcommand{\ist}{\hspace*{.3mm}}
\newcommand{\rmv}{\hspace*{-.3mm}}
\newcommand{\iist}{\hspace*{1mm}}
\newcommand{\rrmv}{\hspace*{-1mm}}
\newcommand{\nn}{\nonumber}
\newcommand{\T}{\mathrm{T}}
\newcommand{\CH}{\mathrm{H}}

\newcommand{\C}{\mathbb{C}}

\definecolor{myblue}{RGB}{79, 129, 189}
\definecolor{myorange}{RGB}{247, 150, 70}
\definecolor{IEEEblue}{RGB}{0 98 155}
\colorlet{blue}{IEEEblue}

\DeclareMathAlphabet{\mathpzc}{OT1}{pzc}{m}{it}
\DeclareMathOperator{\sdsum}{\sum \rmv \rmv \cdots \rmv \rmv \sum}
\DeclareMathOperator{\idint}{\int \rmv \rmv \cdots \rmv \rmv \int}

\begin{document}
\title{AI-enhanced Direct SLAM:\\\LARGE{A Principled Approach to Unsupervised Learning in Bayesian Inference}
	%Unsupervised Learning for AI-augmented Bayesian Inference: Direct Localization and Mapping\vspace*{-1mm}
%Neural-Enhanced Direct Localization and Mapping
}
\author{\IEEEauthorblockN{Alexander~Venus$^*$,~%~\IEEEmembership{Member,~IEEE},
	Benjamin~Deutschmann$^*$,~%~\IEEEmembership{Student Member,~IEEE},
	Alexander Fuchs$^+$,
	Christian Knoll$^+$,
	and~Erik~Leitinger$^*$ %,~\IEEEmembership{Member,~IEEE} 
	}\\[1mm]
	\IEEEauthorblockA{
	$^*$Institute of Comm. Networks and Satellite Comms., Graz University of Technology, Austria, 	$^+$Levata GmbH\\
	{\small Email: \{a.venus,benjamin.deutschmann,erik.leitinger\}@tugraz.at, \{fuchs,knoll\}@levata.at} 
	\thanks{The project has received funding in part from the Austrian Research Promotion Agency (FFG) under the PRISM project (620753) and in part by the European Union’s Horizon Europe research and innovation program under Grant 101192113.}
	}
\vspace*{-9mm}
}

\maketitle
\renewcommand{\baselinestretch}{0.99}\small\normalsize
\begin{abstract}	
	%Multipath propagation, often detrimental to conventional positioning, provides valuable geometric information for \ac{slam} in challenging 6G environments with obstructed line-of-sight. Yet, existing multipath-based direct SLAM methods rely on simplified parametric environment models that cannot capture diffuse and complex scattering, leading to model mismatch.
	
	In this paper, we propose an \ac{ai}-enhanced hybrid \ac{slam} method that performs Bayesian inference directly on raw \ac{rf} signals while learning an environment model in an unsupervised manner. The approach combines a physically interpretable signal model for \ac{los} components with an \ac{ai} model that captures multipath component statistics.
	Building on this formulation, we develop a particle-based \ac{spa} on a factor graph that jointly estimates the \ac{mt} state, visibility, multipath parameters, and noise variances, and integrate it into a variational framework that maximizes the \ac{elbo} to learn the \ac{nn} parametrization directly from measurements. We further present a highly efficient GPU-based implementation that enables parallel likelihood evaluation across particles and \acp{bs}. Simulation results in multipath environments demonstrate that the proposed method learns the generative, environment-dependent signal model in an unsupervised manner while accurately localizing the \ac{mt} and effectively exploiting the learned map in \ac{olos} scenarios.
\end{abstract}
\renewcommand{\baselinestretch}{0.95}\small\normalsize

\IEEEpeerreviewmaketitle

\section{Introduction}\label{sec:intro}
Accurate localization of \acfp{mt} and situational awareness of the surrounding propagation environment are key enablers for future 6G communication networks, supporting applications such as autonomous navigation, indoor localization, digital twinning, and integrated sensing and communications \cite{GonFurKalValDarSheSheBayWymProcIEEE2024}. While conventional positioning techniques perform well in open environments, their reliability degrades in urban and industrial scenarios as well as indoor environments due to multipath propagation and \acl{olos} conditions. Yet, multipath propagation also carries rich geometric information that can be exploited for localization \cite{WitMeiLei:J16, GenJosWan:J16}.

\Acf{mpslam} exploits reflected \acp{mpc} from map features such as \acp{va}, point scatterers, or simple surface models to jointly infer the \ac{mt} state and environment from estimated \ac{mpc} parameters (such as \ac{toa}, \ac{doa}, \ac{dod}, or Doppler) \cite{HanFleuRao:TSP2018, GreLeiWitFle:J24} within a Bayesian framework, and is often solved through \ac{bp} on a factor graph \cite{LeiMeyHlaWitTufWin:J19,LeiVenTeaMey:TSP2023,LiaLeiMey:TSP2025} or \ac{rfs} statistics \cite{KimGranSveKimWym:TVT2022, GeKalXiaGarAngKimTalValWymSev:TSP2025}. However, commonly used parametric models compress the true propagation environment into a few geometric features and thus fail to capture fine-grained electromagnetic effects such as diffuse multipath, material-dependent scattering, extended structures, as well as hardware imperfections. The same limitation persists in direct \ac{mpslam} \cite{LiaLeiMey:TSP2025}: although the likelihood is evaluated on raw \ac{rf} signals, the underlying generative model remains overly idealized, leading to model mismatch and calls for learnable environment representations.

Recent advances in \ac{ai} demonstrate that purely data-driven models can enable \ac{rf} signal-based localization \cite{SalRupSch:TWC2024, PanHuaCheZhaHaeWym:Arxiv2025}, yet they require large labeled data sets or often generalize poorly under domain shifts. This has motivated hybrid methods that retain physically interpretable models while learning only components that are difficult to model analytically \cite{ShlFarEldGol:TSP2022, ShlWhaEldDim:JRPOC2023, VenLeiTerWit:JSP2023, MerRevShlRouRuu:JVT2024, ChaCorCruMag:Arxiv2025, WeiLiaMey:TSP2026}. For environment learning and mapping, a promising training objective is the marginal likelihood (evidence) of the received data under a latent-variable generative model. Since direct evidence maximization is intractable, variational inference maximizes an \ac{elbo} \cite{TzikasRTSP2008}, enabling principled unsupervised learning via variational methods \cite{JohDuvWilAdaDat:NIPS2016, GreSteSch:NeuRIPS2017}. Unlike supervised localization networks, \ac{elbo}-based learning ``probabilistically matches'' the learned generative model and the observed data.

In this work, we propose an \emph{\ac{ai}-enhanced \ac{slam}} method that tightly couples Bayesian inference with unsupervised environment learning, eliminating the need for large sets of labeled \ac{mt} positions and explicit geometric map information. Therefore, we consider a measurement likelihood function that combines model-based components (capturing the \ac{los} contribution) with learned \ac{ai} components (capturing the environment-induced \acp{mpc}).
This allows us to retain a physically consistent signal model that explicitly models the sampled \ac{rf} signal spectrum and array response and that learns the environment in terms of amplitude statistics and propagation geometry. 
The proper combination of both components is then enforced by a joint statistical model, expressed as type-II likelihood function \cite{HanFleuRao:TSP2018, GreLeiWitFle:J24, LiaLeiMey:TSP2025}, with its covariance matrix fusing model-based and learned contributions. We formulate the problem as a Bayesian inference task on a factor graph that directly takes the raw RF signals as input.
Consequently, we can use particle-based \ac{spa} as an effective framework for joint optimization, i.e., analytical inference of the posterior \acs{pdf} in combination with principled unsupervised learning of the \ac{ai} components by maximization of the data evidence via the \ac{elbo}. Besides its conceptual elegance, this joint hybrid model on top of the proposed factor graph admits a computationally efficient GPU-based implementation with parallelization across particles and \acp{bs}. 
%The approach is formulated as Bayesian inference on a factor graph, where a particle-based \ac{spa} operates directly on raw \ac{rf} signals. The measurement likelihood function combines model-based components with \ac{ai} models that capture the \ac{los} contribution and environment-induced \acp{mpc}.
%The \ac{ai} models are learned in a principled manner by maximizing the data evidence via the \ac{elbo}, using the posterior \acp{pdf} provided by the inference algorithm, which enables fully unsupervised learning directly from \ac{rf} signals. We retain a physically interpretable signal model by explicitly modeling the sampled signal spectrum and array response and employ \acp{nn} only to model environment-dependent \ac{mpc} statistics, in particular amplitude statistics and propagation geometry. The resulting statistical model is expressed as a type-II likelihood function \cite{HanFleuRao:TSP2018, GreLeiWitFle:J24, LiaLeiMey:TSP2025} whose covariance matrix fuses model-based and learned components and admits an computationally efficient GPU-based evaluation, enabling parallel processing across particles and \acp{bs}. 
The main contributions are summarized as follows.
\begin{itemize}
	\item We propose an \ac{ai}-enhanced \ac{slam} method that combines a physically interpretable signal model with \ac{nn} models.
	\item We develop a factor graph-based particle \ac{spa} with computationally efficient covariance evaluation for scalable inference directly on raw \ac{rf} measurements.
	\item We introduce an unsupervised learning approach that maximizes the \ac{elbo} to learn the propagation geometry and amplitude statistics directly from \ac{rf} signals.
	\item We present a efficient GPU-based implementation enabling parallel likelihood evaluation across particles and \acp{bs}.
\end{itemize}

\section{System Model and Problem Formulation} \label{sec:model}
%An \ac{mt} at unknown position $\V{p}_k \in \mathbb{R}^d$ transmits a signal at center frequency $f_\mathrm{c}$ and bandwidth $B$. In the frequency domain, the transmitted signal in baseband is $S(f) \rmv\in\rmv \mathbb{C}$. At each time step $k$, $M$ samples of the received signal waveform are recorded by the \acp{bs} at known position $\V{p}^{(j)} \in \mathbb{R}^d$. In the frequency domain, the resulting $M_\text{f} = B/\Delta + 1$ samples have a frequency spacing of $\Delta$. 
Each \ac{bs} $j$ located at known position $\V{p}^{(j)} \in \mathbb{R}^d$ transmits a signal with center frequency $f_\mathrm{c}$ and bandwidth $B$, which is received by a single \ac{mt} at unknown position $\V{p}_k \in \mathbb{R}^d$. In frequency domain, the transmitted baseband signal is denoted by $S(f) \in \mathbb{C}$ with $\|S(f)\|^2 = 1$. At each time step $k$, $M$ samples of the received signal waveform are recorded at the \ac{mt}. This results in $M_\text{f} = B/\Delta + 1$ samples with frequency spacing $\Delta$ (the unambiguous observation distance is given by $d_\mathrm{max} = c/\Delta$, where $c$ is the speed of light). 
%
%The corresponding maximum unambiguous observation distance is given by $d_\mathrm{max} = c/\Delta$, where $c$ is the speed of light. 
The \ac{mt} is equipped with an antenna array comprising $M_\text{a}$ elements, located at positions $\V{a}_{m} \in \mathbb{R}^d$ with $m \in \{1,\ldots,M_\text{a}\}$ (within the frame of reference of the \ac{mt}). The array center is defined to be $\frac{1}{M_\text{a}}\sum_{m=1}^{M_\text{a}} \V{a}_{m} = \V{0}$. In the global coordinate system at time~$k$, the array orientation is denoted by $o_{k}$ and the array center is at $\V{p}_k$.

The \ac{rf} signal model for the sampled received signal with respect to \ac{bs} $j$, i.e.,  $\V{z}^{(j)}_{k} = \big[z_k^{(j)}(-(M-1)/2 \ist \Delta) \hspace{1mm} \cdots \hspace{1mm} z_k^{(j)} ( (M-1)/2 \ist \Delta )\big]^\T = \big[z_{k,1}^{(j)} \hspace{1mm} \cdots \hspace{1mm} z_{k,M}^{(j)} \big]^\T\rmv\rmv\rmv$ comprising the \ac{los} component and a superposition of $L_k^{(j)}-1$ \acp{mpc} can be expressed as \cite{LiaLeiMey:TSP2025}
\vspace*{-4mm}
\begin{align}
	\V{z}^{(j)}_k
	=
	\sum\nolimits_{l=1}^{L_k^{(j)}}
	\varrho_{l,k}^{(j)}\,
	\V{h}_{\V{\chi}}\big(\tau_{l,k}^{(j)},\V{u}_{l,k}^{(j)}\big)
	+
	\V{\epsilon}_k^{(j)}
	\label{eq:radiosigmod}\\[-7mm]\nn
\end{align}
where $\tau_{l,k}^{(j)}$ is the delay, $\V{u}_{l,k}^{(j)}$ is the \ac{doa}, and $\varrho_{l,k}^{(j)}\in\mathbb{C}$ is the complex amplitude of the $l$-th MPC and $\V{\epsilon}_k^{(j)}\sim\mathcal{CN}(\V{0},\eta_k^{(j)}\M{I})$ is
additive noise. Assuming a planar-wave, narrow-band signal model \cite{RichterPhD2005}, the according joint frequency--array response vector $\V{h}_{\V{\chi}}(\tau,\V{u})$ with length $M = M_\text{f}M_\text{a}$ related to the parameter \ac{toa} $\tau$ and \ac{doa} $\V{u}$ of an \ac{mpc} factorizes as 
\vspace*{-1mm}
\begin{align}
	\V{h}_{\V{\chi}}(\tau,\V{u})
	=
	\V{h}_{\V{\chi}_\text{f}}(\tau)\otimes \V{a}_{\V{\chi}_\text{u}}(\V{u}) \in \mathbb{C}^{M}\label{eq:arrayres}\\[-6mm]\nn
\end{align}
where the vector $\V{\chi} = [\V{\chi}_\text{f}^\T \ist \ist \V{\chi}_\text{u}^\T]^\T$ contains unknown calibration parameters of the response vector and $\V{h}_{\V{\chi}_\text{f}}(t)\in\mathbb{C}^M_\text{f}$ denotes the sampled transmitted signal delayed by the \ac{toa} $\tau$, i.e.,
\vspace*{-1mm}
\begin{align}
	\V{h}_{\V{\chi}_\text{f}}(\tau) &=  \frac{c}{4\pi f_\text{c}\tau}\big[w_{\text{f},1} S \big( -(M_\text{f}-1)/2 \ist \Delta \big) \ist\mathrm{e}^{j2\pi (M_\text{f}-1)/(2 \ist \Delta) \tau} \nn\\[-.5mm] & \cdots \hspace{1mm} w_{\text{f},M_\text{f}} S \big( (M_\text{f}-1)/2 \ist \Delta \big) \ist\mathrm{e}^{-j2\pi (M_\text{f}-1)/(2 \ist \Delta) \tau}\big]^\T \rmv\rmv 
	\label{eq:ref_signal}\\[-6mm]\nn
\end{align} 
where $c/(4\pi f_\text{c}\tau)$ accounts for the path-loss and $\V{\chi}_\text{f} = [w_{\text{f},1} \ist \cdots\ist w_{\text{f},M_\text{f}}]^\T$ accounts for mismatches in the transmitted baseband signal and receiver matched filters. The proposed method can also be extended to unsynchronized systems \cite{GenJosWan:J16, KimGranSveKimWym:TVT2022, FasDeuKesWilColWitLeiSecWym:STSP2025}. The array response $\V{a}_{\V{\chi}_\text{u}}^{(j)}(\V{u})$ for direction $\V{u} \in\mathbb{R}^d$ with $\|\V{u}\|=1$ is given by
\vspace*{-1mm}
\begin{align}
	\V{a}_{\V{\chi}_\text{u}}(\V{u})
	&=
	\big[
	w_{\text{u},1}\ist e^{-j\frac{2\pi}{\gamma_c}\V{u}^\T \big(\V{a}_{1}+\Delta\V{a}_{1}\big)}\hspace{1mm} \nn\\[-1mm] &\hspace{14mm}\cdots \hspace{1mm} 
	w_{\text{u},M_\text{a}}\ist e^{-j\frac{2\pi}{\gamma_c}\V{u}^\T\big(\V{a}_{M_\text{a}}+\Delta\V{a}_{M_\text{a}}\big)}
	\big]^\T\\[-7mm]\nn
\end{align}
where ${\V{\chi}_\text{u}} = [w_{\text{u},1} \ist \cdots\ist w_{\text{u},M_\text{a}} \iist \Delta\V{a}^{\T}_{1} \ist \cdots\ist \Delta\V{a}^{\T}_{M_\text{a}}]^\T$ accounts for antenna array mismatches. The geometry dimension $d$ is specified by the specific scenario ($2$D in this work).

\subsection{\Ac{ai}-augmented Measurement Model and System Model}
\label{subsec:meas_model}

%The \ac{mt} state at time $k$ is denoted by $\V{x}_k$ and comprises the \ac{mt} position $\V{p}_k$ along with relevant motion parameters. 
By separating the \ac{los} component from the \ac{mpc} component and explicitly combining the frequency-array response vector in \eqref{eq:arrayres} with \ac{ai}-based models for the \ac{mpc} contribution, the signal model in \eqref{eq:radiosigmod}, conditioned on the \ac{mt} position $\V{p}_k$ and orientation $o_k$, can be rewritten as\footnote{The proposed signal model can be straightforwardly extended to \acs{mimo} systems or adapted to include Doppler frequencies.}
\vspace*{-1mm}
\begin{align}
	\hspace*{-2mm}\V{z}_k^{(j)}
	&\rmv=\rmv
	r_k^{(j)}\rho_{\text{\scriptsize lo},k}^{(j)}\,\V{h}_{\V{\chi}}^{\text{\scriptsize lo}\ist(j)}(\V{p}_k)
	\rmv+\rmv\rmv\sum\nolimits_{n=1}^{D}\rho^{\text{\scriptsize ai}\ist(j)}_{\V{\theta},n}\,\V{h}^{\text{\scriptsize ai}\ist(j)}_{\tilde{\V{\theta}},n}(\V{p}_k)
	\rmv+\rmv\V{\epsilon}_k^{(j)}\hspace*{-2mm}
	\label{eq:measmod}\\[-6mm]\nn
\end{align}
where $\V{h}_{\V{\chi}}^{\text{\scriptsize lo}\ist(j)}(\V{p}_k) \triangleq \V{h}_{\V{\chi}}\big(\tau_{\text{\scriptsize lo},k}^{(j)}, \V{u}_{\text{\scriptsize lo},k}^{(j)}\big) \in \mathbb{C}^{M}$ denotes the signal contribution from the \ac{los} of \ac{bs} $j$. This term is parameterized by the delay $\tau_{\text{\scriptsize lo},k}^{(j)} = \|\V{p}^{(j)}-\V{p}_k\|/c$ and the \ac{doa} $\V{u}_{\text{\scriptsize lo},k}^{(j)} = \M{R}(o_k)(\V{p}^{(j)}-\V{p}_k)/\|\V{p}^{(j)}-\V{p}_k\|$, where the rotation matrix $\M{R}(o_k)$ transforms the direction from the \ac{mt} local coordinate system to the global coordinate system according to the \ac{mt} orientation $o_k$.
%The \ac{mt} state at time $k$ is denoted by $\V{x}_k$ and includes the \ac{mt} position $\V{p}_k$ and possibly further motion-related variables. Separating the \ac{los} and \ac{mpc} contributions that represent the propagation environment, the signal model \eqref{eq:radiosigmod} can be rewritten as\footnote{The proposed signal model for \ac{ne} inference can be straightforwardly extended to \acs{mimo} systems or consider Doppler frequencies.}
%\begin{align}
%	\hspace*{-1mm}\V{z}_k^{(j)}
%	&\rmv=\rmv
%	r_k^{(j)}\rho_{\text{\scriptsize lo},k}^{(j)}\,\V{h}_{\V{\chi}}^{\text{\scriptsize lo}\ist(j)}(\V{p}_k)
%	\rmv+\rmv\rmv\sum_{n=1}^{D}\rho^{\text{\scriptsize ai}\ist(j)}_{\V{\theta},n,k}\,\V{h}^{\text{\scriptsize ai}\ist(j)}_{\tilde{\V{\theta}},n}(\V{p}_k)
%	\rmv+\rmv\V{\epsilon}_k^{(j)}
%	\label{eq:measmod}
%\end{align}
%where $ \V{h}_{\V{\chi}}^{\text{\scriptsize lo}\ist(j)}(\V{p}_k) \triangleq \V{h}_{\V{\chi}}\big(\tau_{\text{\scriptsize lo},k}^{(j)},
%\V{u}_{\text{\scriptsize lo},k}^{(j)}\big) \in \mathbb{C}^{M}$ denotes the contribution related to the \ac{los} of \ac{bs} $j$ parameterized by the delay $\tau_{\text{\scriptsize lo},k}^{(j)} = \|\V{p}_k-\V{p}^{(j)}\|/c$ and \ac{doa} $\V{u}_{\text{\scriptsize lo},k}^{(j)} = \M{R}(o_k)(\V{p}_k-\V{p}^{(j)})/\|\V{p}_k-\V{p}^{(j)}\|$, where $\M{R}(\cdot)$ is a rotation matrix that rotates the direction from the \ac{mt} coordinate system to the global coordinate system considering the \ac{mt} orientation $o_k$. 
The response vector accounting for \ac{ne} components is given by $\V{h}^{\text{\scriptsize ai}\ist(j)}_{\tilde{\V{\theta}},n}(\V{p}_k) \triangleq \V{h}_{\V{\chi}}\big(\tau^{\text{\scriptsize ai}\ist(j)}_{\V{\theta},n,k},\V{u}^{\text{\scriptsize ai}\ist(j)}_{\V{\theta},n,k}\big) \in \mathbb{C}^{M}$, where $\tilde{\V{\theta}} = [\V{\chi}^\T \ist \V{\theta}^\T]^\T$ collects the unknown calibration parameters and the \ac{ai}-model parametrization.  The \ac{ne} components account for the contribution vector related to \ac{mpc} originating from the surrounding propagation environment with parameters $\V{\tau}^{\text{\scriptsize ai}\ist(j)}_{\V{\theta},n,k} = \|\V{p}^{\text{\scriptsize ai}\ist(j)}_{\V{\theta},n} - \V{p}_k\|/c + b^{\text{\scriptsize ai}\ist(j)}_{\V{\theta},n}$ and $\V{u}^{\text{\scriptsize ai}\ist(j)}_{\V{\theta},n,k} = \M{R}(o_k)(\V{p}^{\text{\scriptsize ai}\ist(j)}_{\V{\theta},n} - \V{p}_k)/\|\V{p}^{\text{\scriptsize ai}\ist(j)}_{\V{\theta},n} - \V{p}_k\|$ with neural map feature positions $\V{p}^{\text{\scriptsize ai}\ist(j)}_{\V{\theta},n}$ $\in \mathbb{R}^{d \times 1}$ and biases ${b}^{\text{\scriptsize ai}\ist(j)}_{\V{\theta},n}$ $\in \mathbb{R}$ describing the geometry of the propagation environment represented by data-driven \ac{ml} model given by
\begin{align}
	\V{P}^{\text{\scriptsize ai}\ist(j)}_{\V{\theta}} &=  f^{\text{\scriptsize ai}}_{\V{p},\V{\theta}}( \V{p}^{(j)}) \iist \in \mathbb{R}^{d+1 \times D}\label{eq:nnpos}
\end{align}
where $\V{P}^{\text{\scriptsize ai}\ist(j)}_{\V{\theta}} \!=\! \big[[\V{p}^{\text{\scriptsize ai}\ist(j)\ist \T}_{\V{\theta},1} {b}^{\text{\scriptsize ai}\ist(j)}_{\V{\theta},1}]^\T \ist \cdots \ist [\V{p}^{\text{\scriptsize ai}\ist(j)\ist \T}_{\V{\theta},D} {b}^{\text{\scriptsize ai}\ist(j)}_{\V{\theta},D}]^\T\big]$. Note that \eqref{eq:nnpos} does not depend on the \ac{mt} position, since it represents a \ac{mt} position invariant map. The number of components $D$ of the \ac{ai} model is a design parameter discussed later in Section~\ref{sec:data_model}.
The existence of the \ac{los} component of each \ac{bs} $j$ is modeled by a binary random variable $r_{k}^{(j)} \in \{0, 1\}$. Thus, the direct path between \ac{mt} and \ac{bs} $j$ exists and contributes to the received signal, if and only if $r_{k}^{(j)} \rmv=\rmv 1$. 	

We assume that the complex amplitudes, $\rho_{\text{\scriptsize lo,}k}^{(j)}$ and $\rho^{\text{\scriptsize ai}\ist(j)}_{\V{\theta},n}$, are random and unknown \cite{LiaLeiMey:TSP2025,WipRao:J07, BadHanThoFle:J17}. In particular, random amplitudes are modeled by zero-mean complex Gaussian random variables $\rho_{\text{\scriptsize lo},k}^{(j)} \sim \mathcal{CN}\big(\rho_{\text{\scriptsize lo},k}^{(j)}; 0,\gamma^{(j)}_{k}\big)$ with amplitude variance $\gamma^{(j)}_{k}$ and $\rho^{\text{\scriptsize ai}\ist(j)}_{\V{\theta},n,k} \sim \mathcal{CN}\big(\rho^{\text{\scriptsize ai}\ist(j)}_{\V{\theta},n}; 0,\gamma^{\text{\scriptsize ai}\ist(j)}_{\V{\theta},n}\big)$ with variance 
\begin{equation} \label{eq:nnrho}
	\V{\gamma}^{\text{\scriptsize ai}\ist(j)}_{\V{\theta}} =  f^{\text{\scriptsize ai}}_{\rho,\V{\theta}}(\V{p}^{(j)}) \in \mathbb{R}^{1 \times D}_+ \ist 
\end{equation}
represented by a data-driven \ac{ml} model with $\V{\gamma}^{\text{\scriptsize ai}\ist(j)}_{\V{\theta}} = [\gamma^{\text{\scriptsize ai}\ist(j)}_{\V{\theta},1}\ist \cdots \ist\gamma^{\text{\scriptsize ai}\ist(j)}_{\V{\theta},D}]^\T$. 
%Note that the amplitude model is assumed \ac{mt}-position invariant due to path-loss compensation. To capture position-dependent variations caused by complex environment interactions and visibility changes of feature-related propagation paths, an additional position-dependent network can be incorporated.
The amplitude model\footnote{In the Swerling-1 model, the target reflectivity is modeled as a zero-mean circularly symmetric complex Gaussian random variable, which implies a uniformly distributed phase and a Rayleigh-distributed magnitude; by deliberately neglecting phase information and mean amplitude and characterizing the return solely through its variance (average power), the model is particularly robust against both amplitude and phase fluctuations.} is referred to as \textit{Swerling} 1 \cite{LepRabLeG:TAES2016}. The complex amplitudes $\rho_{\text{\scriptsize lo,}k}^{(j)}$ and $\rho_{{\text{\scriptsize ai}},n}^{(j)}$ are independent across $k, n$, and $j$. 
The additive noise term $\V{\epsilon}^{(j)}_{k}\rmv\rmv$ is distributed according to $\mathcal{CN}\big(\V{\epsilon}^{(j)}_{k}; \V{0}, \eta_{k}^{(j)} \M{I}_M \big)$ and independent across $k$ and $j$ with noise variance $\eta_{k}^{(j)}$. 

From the signal model in \eqref{eq:measmod}, we obtain the \ac{pdf} of the measurement $\V{z}^{(j)}_{k}$ conditioned on the \ac{mt} state $\V{x}_{k} = [\V{p}^\T_k \ist \V{v}^\T_k \ist o_k]^\T$ (containing respectively its position, velocity, and orientation), the \ac{los} state $\V{y}^{(j)}_{k} = [r_{k}^{(j)}\ist \gamma_{k}^{(j)}]^\T$, and the noise variance $\eta_{k}^{(j)}$ to be zero-mean complex Gaussian \ac{pdf} and given as 
\begin{equation}
	\hspace*{-2mm}f_{\tilde{\V{\theta}}}(\V{z}_{k}^{(j)} | \V{x}_k, \V{y}_{k}^{(j)}\rmv\rmv, \eta_{k}^{(j)}) \rmv\rmv=\rmv\rmv \mathcal{CN}\big(\V{z}_{k}^{(j)}; \V{0}, \M{C}_{\tilde{\V{\theta}},k}^{(j)}(\V{p}_k,\V{y}_{k}^{(j)}\rmv\rmv,\eta_{k}^{(j)})\big) \label{eq:likelihood}
\end{equation}
with covariance matrix
\begin{align}
	\M{C}^{(j)}_{\tilde{\V{\theta}},k}(\V{p}_k,\V{y}_{k}^{(j)}\rmv\rmv,\eta_{k}^{(j)}) &=\rmv \eta_{k}^{(j)} \M{I}_M +  r_{k}^{(j)} \M{C}^{(j)}_{\V{\chi},k}(\V{p}_k,\gamma_{k}^{(j)}) 
	\nn\\
	&\hspace*{5mm}+ 
	\M{C}^{\text{\scriptsize ai}\ist(j)}_{\tilde{\V{\theta}},k}(\V{p}_k)\label{eq:fullCov}
\end{align}
where
\begin{align}
	&\hspace*{-2mm}\M{C}^{(j)}_{\V{\chi},k}\rmv(\V{p}_k,\gamma_{k}^{(j)}) = \gamma_{k}^{(j)} \V{h}_{\V{\chi}}^{\text{\scriptsize lo}\ist(j)}(\V{p}_k) \big(\V{h}_{\V{\chi}}^{\text{\scriptsize lo}\ist(j)}\rmv(\V{p}_k)\big)^\CH \label{eq:losCov}\\
	&\hspace*{-2mm}\M{C}^{\text{\scriptsize ai}\ist(j)}_{\tilde{\V{\theta}},k}(\V{p}_k) \rmv\rmv=\rmv\rmv
	\M{P}^{(j)}_{\V{\chi},k}\rmv\rmv
	\sum^{D}_{n = 1} \rmv\rmv\gamma^{\text{\scriptsize ai}\ist(j)}_{\V{\theta},n} \V{h}^{\text{\scriptsize ai}\ist(j)}_{\tilde{\V{\theta}},n}\rmv(\V{p}_k)  \big(\V{h}^{\text{\scriptsize ai}\ist(j)}_{\tilde{\V{\theta}},n}\rmv(\V{p}_k)\big)^\CH
	\M{P}^{(j)\ist \CH}_{\V{\chi},k}\hspace*{-1mm}
	\label{eq:nlos_projection}\\[-7mm]\nn
\end{align}
with $\M{P}^{(j)}_{\V{\chi},k}=\M{I}-\big(\V{h}_{\V{\chi}}^{\text{\scriptsize lo}\ist(j)}(\V{p}_k)\V{h}_{\V{\chi}}^{\text{\scriptsize lo}\ist(j)\ist\CH}(\V{p}_k)\big)/\|\V{h}_{\V{\chi}}^{\text{\scriptsize lo}\ist(j)}(\V{p}_k)\|^2$ being the orthogonal projection operator that enforces that the data-driven covariance matrix is not reducing the position information of the \ac{los} covariance matrix.
Finally, we have $f(\V{z}_{k} | \V{x}_k, \V{y}_{k}) = \prod_{j = 1}^{J}$ $f(\V{z}_{k}^{(j)} |$ $\V{x}_k, \V{y}_{k}^{(j)}, \eta_{k}^{(j)})$, where we introduced $\V{z}_{k} = \big[\V{z}_{k}^{(1) \T} \cdots \V{z}_{k}^{(J) \T}\big]^\T\rmv\rmv\rmv$, $\V{y}_{k} = \big[\V{y}_{k}^{(1) \T} \cdots \V{y}_{k}^{(J) \T}\big]^\T\rmv\rmv\rmv$. Similarly, we also define $\V{\gamma}_{k} = \big[\gamma_{k}^{(1) \T} \cdots \gamma_{k}^{(J) \T}\big]^\T\rmv\rmv\rmv$, $\V{r}_{k} = \big[r_{k}^{(1) \T} \cdots r_{k}^{(J) \T}\big]^\T\rmv\rmv\rmv$, and $\V{\eta}_{k} = \big[\eta_{k}^{(1) \T} \cdots \eta_{k}^{(J) \T}\big]^\T\rmv\rmv\rmv$.

The functional mappings $f^{\text{\scriptsize ai}}_{\V{p},\V{\theta}}(\V{p}^{(j)})$ in \eqref{eq:nnpos} and $f^{\text{\scriptsize ai}}_{\rho,\V{\theta}}(\V{p}^{(j)})$ in \eqref{eq:nnrho} are implemented by deep \acp{nn} as described in Section~\ref{sec:data_model}. These models learn a position-dependent generative \ac{rf} signal model that captures the propagation environment. In addition, the parameter vector $\V{\chi}$ of the response vector in \eqref{eq:arrayres} can be optionally learned in a similar manner (cf. Section~\ref{sec:data_model}).

\subsection{State-Transition Models and Prior PDFs} \label{subsec:state_transition_model}

The dynamics of the \ac{mt} state $\V{x}_{k} = [\V{p}^\T_k \ist \V{v}^\T_k \ist o_k]^\T$ are described by a first-order Markov model with state-transition \ac{pdf} $f(\V{x}_{k}|\V{x}_{k - 1}, \V{z}_o)$, where $\V{z}_o$ is an \ac{imu}-based orientation measurement \cite{LeiWieVenWit:Asilomar2024}. Each \ac{los} state $\V{y}_{k}^{(j)}$, $j \in \{1, \dots, J\}$, and the noise variance are also assumed to evolve independently according to first-order Markov models with transition densities $f(\eta_{k}^{(j)} | \eta_{k - 1}^{(j)})$ and $f(\V{y}_{k}^{(j)} | \V{y}_{k - 1}^{(j)}) = f(\gamma_{k}^{(j)}, r_{k}^{(j)} | \gamma_{k-1}^{(j)}, r_{k-1}^{(j)})$. If the \ac{los} of the \ac{bs}~$j$ is obstructed at time $k-1$, i.e., $r_{k-1}^{(j)} = 0$, it becomes visible at time $k$ with appearance probability $0 < p_{\mathrm{a}} \le 1$ or remains obstructed with probability $1 - p_{\mathrm{a}}$. The corresponding state-transition \ac{pdf} for $r_{k-1}^{(j)} = 0$ is thus given by
\vspace*{-1mm}
\begin{align}
	f(\gamma_{k}^{(j)}, r_{k}^{(j)} | \gamma_{k-1}^{(j)}, 0) = \begin{cases}
		(1-p_\text{a})f_{\mathrm{D}}(\gamma_{k}^{(j)}), & r_{k}^{(j)} = 0 \\
		p_\text{a}f_{\mathrm{a}}(\gamma_{k}^{(j)}), & r_{k}^{(j)} = 1
	\end{cases} \label{eq:state_transition_pf1}\\[-7mm]\nn
\end{align}
where $f_{\mathrm{D}}(\gamma_{k}^{(j)})$ is an arbitrary ``dummy'' \ac{pdf} and $f_{\mathrm{a}}(\gamma_{k}^{(j)})$ is the \ac{pdf} of $\gamma_{k}^{(j)}$ when the \ac{los} of \ac{bs} $j$ appears. If the \ac{bs} is visible at time step $k - 1$, i.e, $r_{k-1}^{(j)} = 1$, then it continues to be visible at time step $k$ with visibility probability $0 < p_{\mathrm{v}} \le 1$. The state-transition \ac{pdf} reads
\begin{align}
	f(\gamma_{k}^{(j)}, r_{k}^{(j)} | \gamma_{k-1}^{(j)}, 1) = \begin{cases}
		(1 - p_{\mathrm{v}}) f_{\mathrm{D}}(\gamma_{k}^{(j)}), &  r_{k}^{(j)} = 0 \\
		p_{\mathrm{v}} f(\gamma_{k}^{(j)} | \gamma_{k-1}^{(j)}), & r_{k}^{(j)} = 1.
	\end{cases}
	\label{eq:state_transition_pf2}\\[-6mm]\nn
\end{align}
If $\gamma_{k}^{(j)}$ is still visible at time $k$, then $\gamma_{k}^{(j)}$ is distributed according to the \ac{pdf}
$f(\gamma_{k}^{(j)} | \gamma_{k - 1}^{(j)} ) = \mathcal{G} ( \gamma_{k}^{(j)}; c_{\gamma}, \gamma_{k - 1}^{(j)} / c_{\gamma} )$ denoting a Gamma\footnote{The Gamma distribution is the conjugate prior for the variance of a complex Gaussian distribution.} \ac{pdf} with mean $\gamma_{k - 1}^{(j)}$\vspace{-.3mm} and variance $(\gamma_{k - 1}^{(j)})^2 / c_{\gamma}$. 
Similarly, the temporal dynamic of the noise variance is described by a \ac{pdf} $f(\eta_{k}^{(j)} | \eta_{k - 1}^{(j)}) = \mathcal{G} ( \eta_{k}^{(j)}; c_{\eta}, \eta_{k - 1}^{(j)} / c_{\eta} )$ with mean $\eta_{k - 1}^{(j)}$\vspace{-.3mm} and variance $(\eta_{k - 1}^{(j)})^2 / c_{\eta}$. At time $k=0$, the prior distributions $f(\V{x}_0)$, $f(\eta_{0}^{(j)})$, $j \in \{1, \dots, J\}$, and $f(\V{y}^{(j)}_{0})$, $j \in \{1, \dots, J\}$ are assumed known. The random variables $\V{x}_0$, $\eta_{0}^{(j)}$, $j \in \{1, \dots, J\}$, and $\V{y}^{(j)}_{0}$, $j \in \{1, \dots, J\}$ are all independent of each other.

\subsection{Declaration and Estimation} \label{subsec:declaration_estimation}

At each time step $k$, given all the measurements $\V{z}_{0 : k} = [\V{z}_{1}^\T \cdots \V{z}_k^\T]^\T\rmv\rmv$, our goal is to estimate \ac{mt} state $\V{x}_{k}$, the noise variance $\eta_{k}^{(j)}$, and \ac{bs} state $\V{y}_{k}^{(j)} $. In our Bayesian setting, the problem comprises the computation of the marginal \acp{pdf} $f(\V{y}_{k}^{(j)}|\V{z}_{1 : k})$, $f(\eta^{(j)}_{k} | \V{z}_{1 : k})$, and $f(\V{x}_{k} | \V{z}_{1 : k})$ using all measurements $\V{z}_{1 : k} \triangleq [\V{z}_{1}^\T \cdots \V{z}_{k}^\T]^\T$. 

Based on these marginal \acp{pdf}, the \ac{mmse} estimate of the \ac{mt} states, \ac{los} states, and noise variances at time $k$, can be obtained as
\vspace*{-1mm}
\begin{align}
	\hat{\V{x}}_{k, \mathrm{MMSE}} &= \int \V{x}_{k} f(\V{x}_{k} | \V{z}_{1 : k}) \hspace{1mm} \mathrm{d} \V{x}_{k} \label{eq:mmse_x} \\[-.5mm]
	\gamma_{k, \mathrm{MMSE}}^{(j)} &= \int \gamma_{k}^{(j)} f(\gamma_{k}^{(j)} | r_{k}^{(j)} = 1, \V{z}_{1 : k}) \hspace{1mm} \mathrm{d} \gamma_{k}^{(j)} \label{eq:mmse_los}\\[-.5mm]
	\hat{\eta}_{k, \mathrm{MMSE}}^{(j)} &= \int \eta_{k}^{(j)} f(\eta_{k}^{(j)} | \V{z}_{1 : k}) \hspace{1mm} \mathrm{d} \eta_{k}^{(j)} \label{eq:mmse_noise} \ist.\\[-7mm]\nn
\end{align}
The marginal \ac{pdf} in \eqref{eq:mmse_los} is obtained by  $f(\gamma_{k}^{(j)} | r_{k}^{(j)} \rmv\rmv=1, \V{z}_{1 : k})$ $ = \rmv\rmv f(\gamma_{k}^{(j)}, r_{k}^{(j)}\rmv\rmv=\rmv\rmv 1|\V{z}_{1 : k})/p(r_{k}^{(j)}\rmv\rmv=\rmv\rmv1  | \V{z}_{1 : k})$, where $p(r_{k}^{(j)} | \V{z}_{1 : k}) =$ $ \int f(\gamma_{k}^{(j)} ,r_{k}^{(j)}| \V{z}_{1 : k}) \mathrm{d} \gamma_{k}^{(j)}$ is the marginal \ac{pmf} representing the \ac{bs} visibility.

\begin{figure}[t]%
	\centering%
	\includegraphics[width = \linewidth
	]{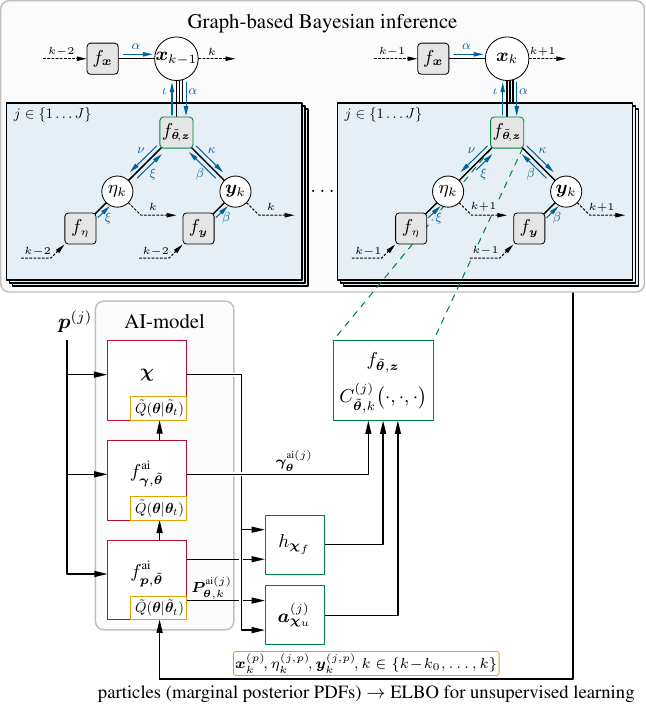}%
	\vspace{-0.2cm}\caption{Concept figure of two time steps of the proposed method, where \acp{nn} learn the environment to enhance the statistical measurement model. Top: factor graph and model-based components (measurement model in green) of the \acf{spa}. Bottom: \ac{ai} model in red and their links to the signal and likelihood model. Quantities associated with unsupervised learning are shown in orange.
	}\vspace{-0.2cm}
	\label{fig:factor-graph}%
\end{figure}

Following the statistical model and assumptions in Sections~\ref{subsec:meas_model} to \ref{subsec:state_transition_model}, the joint posterior \ac{pdf} $f_{\tilde{\V{\theta}}}(\V{x}_{0 : k}, \V{y}_{0 : k}, \V{\eta}_{0 : k} | \V{z}_{1 : k})$ with $\V{x}_{0 : k} \triangleq [\V{x}_0^\T \cdots \V{x}_{k}^\T]^\T$, $\V{y}_{0 : k} \triangleq [\V{y}_0^\T \cdots \V{y}_{k}^\T]^\T\rmv\rmv$, and $\V{\eta}_{0 : k} \triangleq [\V{\eta}_{0}^\T \cdots \V{\eta}_{k}^\T]^\T$  can be factorized as
\vspace*{-1mm}
\begin{align}
	&\hspace*{-3mm}f_{\tilde{\V{\theta}}}(\V{x}_{0 : k}, \V{y}_{0 : k}, \V{\eta}_{0 : k} | \V{z}_{1 : k}) = \frac{f_{\tilde{\V{\theta}}}(\V{x}_{0 : k}, \V{y}_{0 : k}, \V{\eta}_{0 : k} , \V{z}_{1 : k})}{f_{\tilde{\V{\theta}}}(\V{z}_{1 : k})} \label{eq:jointPostPDF1}\\
	&\hspace*{-2mm}\propto f(\V{x}_0) \bigg( \prod^J_{j = 1} f(\V{y}^{(j)}_{0} ) f(\eta^{(j)}_{0} ) \bigg)\rmv \prod_{k^\prime = 1}^{k} f(\V{x}_{k'}|\V{x}_{k' - 1}, \V{z}_o)\rmv  \ist \nn \\[-.3mm]
	&\hspace*{-3mm} \times \prod^J_{j = 1} \rmv\rmv f(\V{y}^{(j)}_{k^\prime} | \V{y}^{(j)}_{k^\prime - 1}) f(\eta_{k'}^{(j)} | \eta_{k' - 1}^{(j)}) \ist f_{\tilde{\V{\theta}}}(\V{z}^{(j)}_{k'} | \V{x}_{k'}, \V{y}^{(j)}_{k'}\rmv,\eta_{k'}^{(j)}) \ist. \hspace*{-1mm}\label{eq:jointPostPDF2}\\[-7mm]\nn
\end{align}
Two time steps ($k$ and $k+1$) of the factor graph corresponding to the factorization in \eqref{eq:jointPostPDF2}, are shown in Fig.~\ref{fig:factor-graph}. This factorization facilitates the development of an efficient method for the computation of marginal posterior \acp{pdf} $f(\V{y}_{k}^{(j)} | \V{z}_{1 : k}) \rmv\approx\rmv \tilde{f}(\V{y}_{k}^{(j)})$,  $f(\eta_{k}^{(j)} | \V{z}_{1 : k}) \rmv\approx\rmv \tilde{f}(\eta_{k}^{(j)})$, and $f(\V{x}_{k} | \V{z}_{1 : k})\rmv\approx\rmv \tilde{f}(\V{x}_{k}) $.

% ---------- - ---------- - ----------
% ---------- - ---------- - ----------

\section{\acf{spa}} \label{sec:bp}

Since direct marginalization of the joint posterior PDF in \eqref{eq:jointPostPDF2} over all time steps $k$ is infeasible, we use \ac{bp} message passing by means of the \ac{spa} \cite{KscFreLoe:01, YedFreWei:05, KolFri:B09} and approximate the marginal posterior \acp{pdf} and messages, typically referred to as beliefs, by a particle-based representation. \ac{bp} is an efficient approach for solving high-dimension Bayesian inference problems. A BP method is derived by performing local operations called ``messages'' on the edges of the factor graph \cite{KscFreLoe:01, YedFreWei:05, KolFri:B09} that represents the statistical model of the Bayesian estimation problem. We make use of the following message passing computation order: (i) messages are passed only forward in time, (ii) along the edges connecting an \ac{mt} state variable node ``$\V{x}_{k}$'' and a \ac{bs}-related variable node ``$\V{y}_{k}^{(j)}$'' or ``$\eta_{k}^{(j)}$'' messages from other \ac{bs}-related variable nodes ``$\V{y}_{k}^{(j')}$'' or ``$\eta_{k}^{(j')}$'' with $j' \neq j$ are not considered (c.f. \cite{LeiMeyHlaWitTufWin:J19}). \ac{bp} messages are also shown in Fig.~\ref{fig:factor-graph}. They are computed in parallel for all \acp{bs} $j \in \{1, \dots, J\}$.

\subsection{Prediction Messages} \label{subsec:bp_pred}

First, we present the \ac{mt} prediction message that is passed from factor node ``$f(\V{x}_k | \V{x}_{k - 1})$'' to variable node ``$\V{x}_k$''. This message is obtained\vspace{-1mm} as
\begin{align}
	\alpha(\V{x}_k) = \int f(\V{x}_k | \V{x}_{k - 1}) \tilde{f}(\V{x}_{k - 1}) \hspace{1mm} \mathrm{d} \V{x}_{k - 1} \label{eq:predAgent}\\[-6mm]\nn
\end{align}
where $\tilde{f}(\V{x}_{k - 1})$ is the belief of the \ac{mt} state at the previous time step $k - 1$. The prediction message of the \ac{bs} states is
\vspace*{-1mm}
\begin{align}
	\beta(\V{y}_{k}^{(j)}) = \sum_{r_{k - 1}^{(j)} \in \{0, 1\} } &\int f(\gamma_{k}^{(j)}, r_{k}^{(j)} | \gamma_{k - 1}^{(j)}, r_{k - 1}^{(j)}) \nn \\[-1mm]
	&\times \tilde{f}(\gamma_{k - 1}^{(j)}, r_{k - 1}^{(j)}) \hspace{1mm} \mathrm{d} \gamma_{k - 1}^{(j)} \label{eq:predPA} \\[-7mm]
	\nn
\end{align}
where $\tilde{f}(\V{y}_{k - 1}^{(j)}) = \tilde{f}(\gamma_{k - 1}^{(j)}, r_{k - 1}^{(j)})$ is the corresponding belief at time step $k - 1$. Inserting \eqref{eq:state_transition_pf1} and \eqref{eq:state_transition_pf2} for $f(\gamma_{k}^{(j)}, r_{k}^{(j)} | \gamma_{k - 1}^{(j)}, 1)$ and $f(\gamma_{k}^{(j)}, r_{k}^{(j)} | \gamma_{k - 1}^{(j)}, 0)$, we obtain for $r_{k}^{(j)} = 1$
\vspace*{-1mm}
\begin{align}
	\beta(\gamma_{k}^{(j)},1) &=  p_{\text{v}}\int f(\gamma_{k}^{(j)} | \gamma_{k - 1}^{(j)}) \tilde{f}(\gamma_{k - 1}^{(j)}, 1)\hspace{1mm} \mathrm{d} \gamma_{k - 1}^{(j)} \nn \\[-1mm]
	&\hspace*{10mm} +  p_{\text{a}} \int f_\text{a}(\gamma_{k}^{(j)}) \tilde{f}(\gamma_{k - 1}^{(j)}, 0)  \mathrm{d} \gamma_{k - 1}^{(j)} \label{eq:predPAone} \\[-8mm]
	\nn
\end{align}
and, for $r_{k}^{(j)} = 0$, we have $\alpha(\gamma_{k}^{(j)},0) = \alpha_{k}^{(j)} f_\text{D}(\gamma_{k}^{(j)}) $ with
\vspace*{-1mm}
\begin{align}
	\beta_{k}^{(j)} &=  (1-p_{\text{v}}) \int \tilde{f}(\gamma_{k - 1}^{(j)}, 1)\hspace{1mm} \mathrm{d} \gamma_{k - 1}^{(j)} \nn \\[-1mm]
	& \hspace*{10mm}+  (1-p_{\text{a}}) \int \tilde{f}(\gamma_{k - 1}^{(j)}, 0)  \mathrm{d} \gamma_{k - 1}^{(j)} \ist.\label{eq:predPAtwo} \\[-8mm]
	\nn
\end{align}

Finally, for each $j \rmv\in\rmv \{1, \dots, J\}$, the prediction message that is sent from the factor node ``$f(\eta_{k}^{(j)} | \eta_{k - 1}^{(j)})$'' to the noise variance node ``$\eta_{k}^{(j)}$'', is given by
\vspace*{-2mm}
\begin{align}
	\xi(\eta_{k}^{(j)}) = \int \hspace{-.5mm} f(\eta_{k}^{(j)} | \eta_{k - 1}^{(j)}) \ist \tilde{f}(\eta_{k - 1}^{(j)}) \hspace{.5mm} \mathrm{d} \eta_{k - 1}^{(j)} \label{eq:bp_xi}\\[-8mm]
	\nn
\end{align}
where $\tilde{f}(\eta_{k - 1}^{(j)})$ is the belief of the noise variance computed at the previous time step $k - 1$. 

\subsection{Measurement Update Messages} \label{subsec:bp_meas}

After prediction messages are obtained, a measurement update is performed where messages that provide the information of measurements $\V{z}_{k}$ are obtained. Specifically, messages sent from factor nodes ``$f(\V{z}_{k}^{(j)} | \V{x}_k, \V{y}_{k}^{(j)})$'' to \ac{mt} state ``$\V{x}_k$'', \ac{bs} states ``$\V{y}_{k}^{(j)}$, $j \rmv\in\rmv \{1, \dots, J\}$ are computed. The messages $\iota(\V{x}_k; \V{z}_k^{(j)} )$, $j \rmv\in\rmv \{1, \dots, J\}$ sent from ``$f(\V{z}_{k}^{(j)} | \V{x}_k, \V{y}_{k}^{(j)}, \eta_{k}^{(j)})$'' to ``$\V{x}_k$'' are obtained as
\vspace{-1mm} 
\begin{align}
	\iota(\V{x}_k; \V{z}_k^{(j)} ) &= \sum_{r_{k}^{(j)} \in \{0, 1\} } \int\rmv\rmv\rmv\rmv\int f(\V{z}_{k}^{(j)} | \V{x}_k, \gamma_{k}^{(j)},r_{k}^{(j)}, \eta_{k}^{(j)})\nn\\[-2mm]
	&\hspace*{10mm}\times\beta(\gamma_{k}^{(j)},r_{k}^{(j)}) \xi(\eta_{k}^{(j)})\hspace{1mm} \mathrm{d} \gamma_{k}^{(j)} \mathrm{d} \eta_{k}^{(j)}\ist. \label{eq:bp_iota} \\[-7mm]
	\nn
\end{align}
Note that the measurements $\V{z}_k^{(j)}$, $j \rmv\in\rmv \{1, \dots, J\}$ are observed and thus fixed. This is indicated by the ``;'' notation in $\iota(\V{x}_k; \V{z}_k^{(j)} )$. 

Next, the messages $\kappa(\V{y}_{k}^{(j)}; \V{z}_{k}^{(j)} )$, $j \rmv\in\rmv \{1, \dots, J\}$ passed from ``$f(\V{z}_{k}^{(j)} | \V{x}_k, \V{y}_{k}^{(j)}, \eta_{k}^{(j)})$'' to \ac{bs} ``$\V{y}_{k}^{(j)}$'' are given by
\vspace*{-1mm}
\begin{align}
	\kappa(\V{y}_{k}^{(j)}; \V{z}_{k}^{(j)} )&=  \int\rmv\rmv\rmv\rmv\int \rmv\rmv f(\V{z}_{k}^{(j)} | \V{x}_k, \gamma_{k}^{(j)},r_{k}^{(j)}, \eta_{k}^{(j)})\nn\\
	& \hspace*{10mm}\times \alpha(\V{x}_k)  \xi(\eta_{k}^{(j)}) \mathrm{d} \V{x}_k \mathrm{d} \eta_{k}^{(j)}\rrmv. \label{eq:bp_kappa}\\[-7mm]\nn
\end{align}

Finally, the messages $\nu(\eta_{k}^{(j)}; \V{z}_k^{(j)} )$, $j \rmv\in\rmv \{1, \dots, J\}$ sent from ``$f(\V{z}_{k}^{(j)} | \V{x}_k, \V{y}_{k}^{(j)}, \eta_{k}^{(j)})$'' to the noise variance nodes ``$\eta_{k}^{(j)}$'' are calculated\vspace{-.5mm} as
\begin{align}
	\nu(\eta_{k}^{(j)}; \V{z}_k^{(j)} ) &= \sum_{r_{k}^{(j)} \in \{0, 1\} } \int \rmv\rmv\rmv\rmv \int f(\V{z}_{k}^{(j)} | \V{x}_k, \gamma_{k}^{(j)},r_{k}^{(j)}, \eta_{k}^{(j)}) \nn \\[-1mm]
	& \hspace*{10mm}\times \beta(\gamma_{k}^{(j)},r_{k}^{(j)})  \alpha(\V{x}_k) \mathrm{d} \gamma_{k}^{(j)}  \mathrm{d} \V{x}_k. \label{eq:bp_{k'}u}\\[-7mm]\nn
\end{align}
These messages provide updated information about the noise variance associated with $\V{z}_k^{(j)}$, while taking into account the knowledge of all \acp{los} states as well as the \ac{mt} state.

With the \ac{bp} messages, the beliefs of \ac{mt} state $\V{x}_k$, \ac{los} states $\V{y}_{k, n}^{(j)}$, $n \in \{1, \dots, N_{k}^{(j)}\}$, $j \rmv\in\rmv \{1, \dots, J\}$ and noise variance $\eta_{k}^{(j)}$, $j \rmv\in\rmv \{1, \dots, J\}$ can be obtained as the product of all messages sent to the corresponding variable node, i.e.,
\vspace*{-1mm}
\begin{align}
	\tilde{f}(\V{x}_k) &\propto \alpha(\V{x}_k) \prod^{J}_{j = 1} {\iota}(\V{x}_k; \V{z}_k^{(j)} ) \label{eq:belief1} \\[.7mm]
	\tilde{f}(\V{y}_{k, n}^{(j)}) &\propto \beta(\V{y}_{k, n}^{(j)}) \ist {\kappa}(\V{y}_{k, n}^{(j)}; \V{z}_k^{(j)} ) \label{eq:belief2}  \\[2.5mm]
	\tilde{f}(\eta_{k}^{(j)}) &\propto  \xi(\eta_{k}^{(j)}) \ist {\nu}(\eta_{k}^{(j)}; \V{z}_k^{(j)} ). \label{eq:belief3} \\[-6mm]
	\nn
\end{align}
Computation of these beliefs involves proper normalization such that they integrate and sum to one. The calculated beliefs can then be used for declaration and state estimation as discussed in\vspace{-2mm} Section~\ref{subsec:declaration_estimation}.

\section{Particle-Based Implementation} \label{sec:particle}
In general, it is not possible to calculate the beliefs and \ac{bp} messages discussed in Section~\ref{sec:bp} in closed form. This is because the integrations in \ac{bp} message passing equations \eqref{eq:predAgent}--\eqref{eq:bp_xi} and \eqref{eq:bp_iota}--\eqref{eq:bp_{k'}u} as well as message multiplication in belief calculations in \eqref{eq:belief1}--\eqref{eq:belief3} cannot be performed analytically.

Hence, in this section, we present a computationally feasible particle-based implementation \cite{AruMasGorCla:02, LeiMeyHlaWitTufWin:J19, LiaLeiMey:TSP2025} of the proposed \ac{bp} method. The beliefs $\tilde{f}(\V{x}_k)$, $\tilde{f}(\V{y}_{k}^{(j)}) = \tilde{f}(\gamma_{k}^{(j)}, r_{k}^{(j)})$, and $\tilde{f}(\eta_{k}^{(j)})$ are represented by weighted particle sets $\{(\V{x}_k^{(p)}, w_{\V{x},k}^{(p)})\}_{p=1}^P$, $\{(\gamma_{k}^{(j,p)}, w_{\V{y},k}^{(j,p)})\}_{p=1}^P$, and $\{(\eta_{k}^{(j,p)}, w_{\eta,k}^{(j,p)})\}_{p=1}^P$, for $j\in\{1,\dots,J\}$.
To obtain linear complexity in $P$, we adopt the stacking approach of \cite{MeyHliHla:J16}, where particles of the incoming \ac{bp} messages for each \ac{mt}--\ac{los}--noise variance tuple are ``stacked'' into a joint state, requiring the same number of particles $P$ for all states; this approach is asymptotically optimal. Note that the \ac{los} weights $w_{\V{y},k}^{(j,p)}$ are not normalized, and their sum $p_k^{(j)} = \sum_{p=1}^P w_{\V{y},k}^{(j,p)} \approx \int \tilde{f}(\gamma_k^{(j)},1)\,\mathrm{d}\gamma_k^{(j)}$ approximates the posterior visibility probability $f(r_k^{(j)} = 1 \mid \V{z}_{1:k})$.

%Hence, in this section, we present a computationally feasible particle-based implementation \cite{AruMasGorCla:02, LeiMeyHlaWitTufWin:J19, LiaLeiMey:TSP2025} of the proposed \ac{bp} method. Here, the beliefs $\tilde{f}(\V{x}_k)$, $\tilde{f}(\V{y}_{k}^{(j)}) =  \tilde{f}(\gamma_{k}^{(j)}, r_{k}^{(j)})$, and $\tilde{f}(\eta_{k}^{(j)})$ are represented by sets of weighted particles $\big\{(\V{x}_k^{(p)}, w_{\V{x}, k}^{(p)})\big\}_{p = 1}^P$, $\big\{(\gamma_{k}^{(j, p)}, w_{\V{y}, k}^{(j, p)})\big\}_{p = 1}^P$, $j \rmv\in\rmv \{1, \dots, J\}$ and $\big\{(\eta_{k}^{(j, p)}, w_{\eta, k}^{(j, p)})\big\}_{p = 1}^{P}$, $j \rmv\in\rmv \{1, \dots, J\}$. To keep the computational complexity linear in the number of particles, we adopted an approach introduced in \cite{MeyHliHla:J16} where the particles representing incoming BP messages for each \ac{mt}-\ac{los}-noise variance pair are stacked into a joint incoming message. This stacking approach requires that the \ac{mt} state, the \ac{los} state, and noise variance are represented by the same number of particles, $P$. Note that this stacking approach is asymptotically optimal \cite{MeyHliHla:J16}. The weights representing \ac{los} states, $w_{\V{y}, k}^{(j, p)}$, do not sum to one, instead we have $p_{k}^{(j)} = \sum_{p = 1}^P w_{\V{y}, k}^{(j, p)} \approx \int \tilde{f}(\gamma_{k}^{(j)}, 1) \hspace{1mm} \mathrm{d} \gamma_{k}^{(j)}$, which approximately equals the posterior visibility probability $f(r_{k}^{(j)} = 1 | \V{z}_{1 : k})$.

The particle-based algorithm at each time step $k$ consists of a prediction/birth stage followed by a measurement update and belief calculation:
\begin{enumerate}
	\item Prediction and birth: Given $\{(\V{x}_{k-1}^{(p)}, w_{\V{x},k-1}^{(p)})\}_{p=1}^P$, $\{(\gamma_{k-1}^{(j,p)}, w_{\V{y},k-1}^{(j,p)})\}_{p=1}^P$, and $\{(\eta_{k-1}^{(j,p)}, w_{\eta,k-1}^{(j,p)})\}_{p=1}^P$, the \ac{mt} state particles $\V{x}_k^{(p)}$ and noise-variance particles $\eta_k^{(j,p)}$ are drawn from their transition models with inherited weights, while for each \ac{bs} $j$ only $P' = P-P_\mathrm{a}$ propagated \ac{los} particles $\gamma_k^{(j,p)}$ are retained from the previous step and the remaining $P_\mathrm{a}$ particles are replaced by birth samples from the appearance density, with weights scaled by the visibility probability $p_\text{v}$ and appearance probability $p_\text{a}$.
	\item Measurement update and belief calculation: For each stacked particle triplet $(\V{x}_k^{(p)}, \gamma_k^{(j,p)}, \eta_k^{(j,p)})$, the covariance $\M{C}^{(j)}_{\tilde{\V{\theta}},k}(\V{p}_k^{(p)}, \gamma_k^{(j,p)}, \eta_k^{(j,p)})$ and corresponding likelihood terms are evaluated, unnormalized importance weights for the BP messages are computed, and normalized to obtain the updated particle beliefs $\tilde{f}(\V{x}_k)$, $\tilde{f}(\gamma_k^{(j)}, r_k^{(j)})$, and $\tilde{f}(\eta_k^{(j)})$ with resampling to prevent particle degeneracy.
\end{enumerate}

An approximation of the \ac{mmse} estimates in \eqref{eq:mmse_x}-\eqref{eq:mmse_noise} at time $k$ for the \ac{mt}, \ac{los}, and variance states is obtained directly from the updated weighted particle sets and is given by
\vspace*{-1mm}
\begin{align}
	\hat{\V{x}}_{k} &= \sum_{p = 1}^P w_{\V{x}, k}^{(p)} \V{x}_{k}^{(p)} \label{eq:MTmmseptcl}\\[-1mm]
	\hat{\gamma}_{k}^{(j)} &= \frac{1}{p_{k}^{(j)}} \sum_{p = 1}^P w_{\V{y}, k}^{(j, p)} \gamma_{k}^{(j, p)} \label{eq:losmmseptcl}\\[-1mm]
	\hat{\eta}_{k}^{(j)} &= \sum_{p = 1}^{P} w_{\eta, k}^{(j, p)} \eta_{k}^{(j, p)}\ist. \label{eq:noisemmseptcl}\\[-7mm]\nn
\end{align}

\subsection{Efficient Calculation of Likelihood Function}
\label{sec:likelihood_evaluation}

When computing the likelihood function in \eqref{eq:likelihood}, the covariance matrix in \eqref{eq:fullCov} must be inverted and its determinant evaluated, both of complexity $\mathcal{O}(M^3)$, for each particle $p$ at each \ac{bs} $j$ when computing the likelihood function in \eqref{eq:likelihood}, a reduction of the computational complexity is essential.

For a given \ac{bs} $j$ and a particle realization of the \ac{mt} position $\V{p}_k$, the covariance matrix in \eqref{eq:fullCov} can be expressed as a low-rank perturbation of a scaled identity matrix. Specifically, we collect the \ac{los} and \ac{ne} components into the matrix
\vspace*{-2mm}
\begin{align}
	\V{U}_k^{(j)}
	&\triangleq
	\big[
	r_k^{(j)}\sqrt{\gamma_k^{(j)}}\,
	\V{h}_{\V{\chi}}^{\text{\scriptsize lo}\ist(j)}(\V{p}_k)
	\;\;
	\sqrt{\lambda^{\text{\scriptsize ai}\ist(j)}_{k,\V{\theta},1}(\V{p}_k)}\,
	\V{h}^{\text{\scriptsize ai}\ist(j)}_{\tilde{\V{\theta}},1}(\V{p}_k)
	\nn\\[-0.5mm]
	&\hspace*{12mm}\cdots\;
	\sqrt{\lambda^{\text{\scriptsize ai}\ist(j)}_{k,\V{\theta},D}(\V{p}_k)}\,
	\V{h}^{\text{\scriptsize ai}\ist(j)}_{\tilde{\V{\theta}},D}(\V{p}_k)
	\big]
	\in \C^{M\times R}
	\label{eq:U_ieee}\\[-7mm]\nn
\end{align}
where $R=D+r_k^{(j)}\le D+1$. Using \eqref{eq:U_ieee}, the covariance matrix in \eqref{eq:fullCov} can be rewritten as
\vspace*{-1mm}
\begin{align}
	\M{C}^{(j)}_{\tilde{\V{\theta}},k}
	=
	\eta_k^{(j)}\M{I}_M
	+
	\V{U}_k^{(j)}\V{U}_k^{(j)\CH}.
	\label{eq:cov_lowrank_ieee}\\[-6mm]\nn
\end{align}
Inspired by \cite{Tipping2003,LepRabLeG:TAES2016,GreLeiWitFle:J24}, we applying the Woodbury matrix inversion lemma~\cite[eq.\,(159)]{Cookbook} and the matrix determinant lemma~\cite[eq.\,(B.1.16)]{Pozrikidis14GridsGraphsNetworks}, respectively, yields
\vspace*{-1mm}
\begin{align}
	\big(\M{C}^{(j)}_{\tilde{\V{\theta}},k}\big)^{-1}
	&=
	(\eta_k^{(j)})^{-1}\M{I}_M
	-
	(\eta_k^{(j)})^{-2}
	\V{U}_k^{(j)}
	\M{G}_k^{(j)-1}
	\V{U}_k^{(j)\CH}
	\label{eq:woodbury_ieee}\\[-6mm]\nn
\end{align}
where $\M{G}_k^{(j)}\triangleq\M{I}_R+(\eta_k^{(j)})^{-1} \V{U}_k^{(j)\CH}\V{U}_k^{(j)}$
and 
\vspace*{-1mm}
\begin{align}
	\log\det\!\big(\M{C}^{(j)}_{\tilde{\V{\theta}},k}\big)
	&=
	M\log(\eta_k^{(j)})
	+
	\log\det\!\big(\M{G}_k^{(j)}\big).
	\label{eq:logdet_ieee}\\[-6.5mm]\nn
\end{align}

Using \eqref{eq:woodbury_ieee}, the quadratic form within the complex Gaussian likelihood function in \eqref{eq:likelihood} simplifies to
\begin{align}
	\hspace*{-1mm}\V{z}_k^{(j)\CH}(\M{C}^{(j)}_{\tilde{\V{\theta}},k}\big)^{-1}
	\V{z}_k^{(j)} \rmv\rmv=\rmv \|\V{q}_k^{(j)}\|^2 \rmv-\rmv \big\|\M{G}_k^{(j)-1/2}\V{B}_k^{(j)\CH}\V{q}_k^{(j)} \big\|^2.
	\label{eq:quad_ieee}
\end{align}
where $	\V{q}_k^{(j)}\triangleq(\eta_k^{(j)})^{-1/2}\V{z}_k^{(j)}$ and $\V{B}_k^{(j)}\triangleq(\eta_k^{(j)})^{-1/2}\V{U}_k^{(j)}$.
Substituting \eqref{eq:quad_ieee} and \eqref{eq:logdet_ieee} into \eqref{eq:likelihood}, the log-likelihood function in \eqref{eq:likelihood} can be rewritten as
\vspace*{-2mm}
\begin{align}
	&\log f_{\tilde{\V{\theta}}}\big(\V{z}_k^{(j)}|\V{x}_k,\V{y}_k^{(j)},\eta_k^{(j)}\big)\nn\\
	&\hspace*{15mm}=
	-\|\V{q}_k^{(j)}\|^2
	+
	\big\|
	\M{G}_k^{(j)-1/2}
	\V{B}_k^{(j)\CH}
	\V{q}_k^{(j)}
	\big\|^2
	\nn\\
	&\hspace*{20mm}
	-\log\det\!\big(\M{G}_k^{(j)}\big)
	-
	M\log(\pi\eta_k^{(j)})
	\label{eq:loglike_ieee}\\[-7mm]\nn
\end{align}
which requires only inversion and determinant computation of the $R\times R$ matrix $\M{G}_k^{(j)}$, with $R=D+r_k^{(j)}\ll M$. 
Hence, the computational complexity per likelihood evaluation is reduced from $\mathcal{O}(M^3)$ to $\mathcal{O}(MR^2+R^3)$.
Notably, $JP$ computations across \acp{bs} and particles can be computed in parallel.

\textit{Remark:}
Since the \ac{los} contribution is rank-one, the two cases $r_k^{(j)}\in\{0,1\}$ differ only by the presence of a single column in \eqref{eq:U_ieee}. This structure enables efficient ``rank-one'' updates when calculating the likelihood function in \eqref{eq:likelihood} for $r_k^{(j)} = 0$ and $r_k^{(j)} = 1$, respectively.

\section{Variational Unsupervised Learning}	\label{sec:em_learning}
We assume that the \ac{mt} states $\V{x}_{0:k}$, the \ac{los} states $\V{y}_{0:k}=\big[\V{r}_{0:k}\; \V{\gamma}_{0:k}\big]^\T$, and noise variances $\V{\eta}_{0:k}$ are latent variables for all \ac{bs} $j\in\{1,2,\dots,J\}$ and time steps $\{0,2,\dots,k\}$. The proposed unsupervised learning algorithm learns the \ac{ai} parametrization $\tilde{\V{\theta}}$ by maximizing the marginal likelihood function, i.e., \textit{evidence}, given all measurements $\V{z}_{1:k}$, i.e.,
\vspace*{-1mm}
\begin{align}
	\tilde{\V{\theta}}^* &= \text{arg max}_{\tilde{\V{\theta}}} \iist \log	f_{\tilde{\V{\theta}}}(\V{z}_{1 : k})\nn\\
	&= \text{arg max}_{\tilde{\V{\theta}}} \log \underset{\V{r}_{0 : k}}{\sdsum}\underset{\V{x}_{0 : k}\ist \V{\gamma}_{0 : k}\ist \V{\eta}_{0 : k}}{\idint} \nn\\[-1mm]
	&\hspace*{1mm}\times f_{\tilde{\V{\theta}}}(\V{x}_{0 : k}, \V{r}_{0 : k}, \V{\gamma}_{0 : k}, \V{\eta}_{0 : k}, \V{z}_{1 : k})\,
	\mathrm{d}\V{x}_{0:k}\mathrm{d}\V{\gamma}_{0 : k}\mathrm{d}\V{\eta}_{0 : k}\ist .
	\label{eq:ml_marginal}\\[-6mm]\nn
\end{align}
Since direct maximization of \eqref{eq:ml_marginal} is intractable due to the high-dimensional integrals and features and the summation over the discrete LOS variables, we resort to an variational-based learning framework. Following \cite{TzikasRTSP2008}, for any auxiliary density $q(\V{x}_{0:k},\V{y}_{0:k},\V{\eta}_{0:k})$, we can decompose the marginal log-likelihood function as
\vspace*{-1mm}
\begin{align}
	\log f_{\tilde{\V{\theta}}}(\V{z}_{1 : k}) &= \mathcal{L}\big(q(\V{x}_{0 : k}, \V{y}_{0 : k}, \V{\eta}_{0:k});\tilde{\V{\theta}}\big) + \text{KL}\big(q(\V{x}_{0 : k}, \V{y}_{0 : k}, \nn\\
	& \hspace*{5mm} \V{\eta}_{0:k}), f_{\tilde{\V{\theta}}}(\V{x}_{0 : k}, \V{y}_{0 : k}, \V{\eta}_{0:k}| \V{z}_{1 : k})\big)\\[-6mm]\nn
\end{align}
where $\text{KL}(\cdot,\cdot)$ is the \ac{kld} and
\vspace*{-1mm}
\begin{align}
	&\mathcal{L}\big(q(\V{x}_{0:k},\V{r}_{0 : k},\V{\gamma}_{0 : k},\V{\eta}_{0 : k});\tilde{\V{\theta}}\big)\nn\\
	&\hspace*{0mm}=
	\underset{\V{r}_{0 : k}}{\sdsum} \underset{\V{x}_{0 : k}\ist \V{\gamma}_{0 : k}\ist \V{\eta}_{0 : k}}{\idint}  q\big(\V{x}_{0:k},\V{r}_{0 : k},\V{\gamma}_{0 : k},\V{\eta}_{0 : k}\big)\nn\\[-0.5mm]
	&\hspace*{0mm}
	\times \log \frac{f_{\tilde{\V{\theta}}}(\V{x}_{0 : k},\V{r}_{0 : k},\V{\gamma}_{0 : k},\V{\eta}_{0 : k}, \V{z}_{1 : k})}
	{q(\V{x}_{0:k},\V{r}_{0 : k},\V{\gamma}_{0 : k},\V{\eta}_{0 : k})}
	\,\mathrm{d}\V{x}_{0:k}\mathrm{d}\V{\gamma}_{0 : k} \mathrm{d}\V{\eta}_{0 : k}\ist.\\[-6mm]\nn
\end{align}
represents the \ac{elbo}. Since the \ac{kld} is non-negative, the \ac{elbo} $\mathcal{L}\big(q(\V{x}_{0:K},\V{r}_{0 : k},\V{\gamma}_{0 : k},\V{\eta}_{0 : k});\tilde{\V{\theta}}\big)$ is a lower bound on the marginal log-likelihood function with equality, if and only if, $q(\V{x}_{0:k},\V{r}_{0 : k},\V{\gamma}_{0 : k},\V{\eta}_{0 : k}) = f_{\tilde{\V{\theta}}}(\V{x}_{0 : k},\V{r}_{0 : k},\V{\gamma}_{0 : k},\V{\eta}_{0 : k}| \V{z}_{1 : k})$ \cite{Bis:B06,TzikasRTSP2008}.

%\subsection{Variational Unsupervised Learning}

Based on this rationale, we formulate a two step iterative algorithm that maximizes the \ac{elbo} using the cost function $q^{(t)}(\V{x}_{0 : k},\V{y}_{0 : k},\V{\eta}_{0 : k}) = f_{\tilde{\V{\theta}}_t}(\V{x}_{0 : k},\V{y}_{0 : k},\V{\eta}_{0 : k}| \V{z}_{1 : k})$, with $\tilde{\V{\theta}}_t$ being the current \ac{ai} parametrization, yielding the auxiliary function
\vspace*{-1mm}
\begin{align}
	&\hspace*{-3mm}Q(\tilde{\V{\theta}}| \tilde{\V{\theta}}_t)\nn\\
	&\hspace*{-3mm}\triangleq
	\mathbb{E}_{f_{\tilde{\V{\theta}}_t}(\V{x}_{0 : k},\V{y}_{0 : k},\V{\eta}_{0 : k}| \V{z}_{1 : k})}
	\!\left[
	\log f_{\tilde{\V{\theta}}}(\V{x}_{0 : k},\V{y}_{0 : k},\V{\eta}_{0 : k}, \V{z}_{1 : k})
	\right]\nn\\
	&\hspace*{-3mm}=\rmv\rmv\rmv\sum_{k'=1}^k \rmv\rmv \sum_{j=1}^J\rmv
	\mathbb{E}_{f_{\tilde{\V{\theta}}_t}(\V{x}_{k'},\V{y}_{k'}^{(j)},\eta^{(j)}|\V{z}_{k'})}
	\!\big[\rmv\rmv
	\log f_{\tilde{\V{\theta}}}\rmv(\V{z}_{k'}^{(j)}| \V{x}_{k'},\V{y}_{k'}^{(j)},\eta_{k'}^{(j)})
	\big]\hspace*{-3mm}
	\label{eq:Q_expected_like}\\[-6mm]\nn
\end{align}
where we have exploited the factorization of the posterior \ac{pdf} in \eqref{eq:jointPostPDF2}.

\subsubsection{Inference Step (\ac{spa} Algorithm)}

The posterior \ac{pdf} $q^{(t)}(\V{x}_{0 : k},\V{y}_{0 : k},\V{\eta}_{0 : k})
= f_{\tilde{\V{\theta}}_t}(\V{x}_{0 : k}, \V{y}_{0 : k}, \V{\eta}_{0 : k}| \V{z}_{1 : k})$ required in \eqref{eq:Q_expected_like} is approximated using the \ac{spa} algorithm of Section~\ref{sec:particle}. Since we perform resampling after each measurement update, we only need the particles of the \ac{mt} state and the variances for each \ac{bs} $\V{x}^{(p)}_{k'}$ and $\eta^{(j,p)}_{k'}$ as well as the particles and existence probabilities of the \ac{los} states $\gamma^{(j,p)}_{k'}$ and $p^{(j)}_{k'}$ to approximate $Q(\tilde{\V{\theta}}| \tilde{\V{\theta}}_t)$ in \eqref{eq:Q_expected_like}, i.e.,
\vspace*{-1mm}
\begin{align}
	\tilde{Q}(\tilde{\V{\theta}}|\tilde{\V{\theta}}_t)
	&=
	\sum_{k'=1}^k\sum_{j=1}^J
	\frac{1}{P}\sum_{p=1}^{P} \Big(p^{(j)}_{k'}
	\log \mathcal{CN}\!\big(
	\V{z}_{k'}^{(j)};\V{0},
	\M{C}_{\tilde{\V{\theta}},n,1}^{(j,p)}
	\big)\nn\\
	&\hspace*{5mm} + \big(1-p^{(j)}_{k'}\big)
	\log \mathcal{CN}\!\big(
	\V{z}_{k'}^{(j)};\V{0},
	\M{C}_{\tilde{\V{\theta}},n,0}^{(j,p)}
	\big)\Big)
	\label{eq:Q_particles}\\
	&= \sum_{k'=1}^k \tilde{Q}_k(\tilde{\V{\theta}}|\tilde{\V{\theta}}_t)\label{eq:Q_particles2}\\[-7mm]\nn
\end{align}
with $\M{C}_{\tilde{\V{\theta}},n,1}^{(j,p)} \triangleq \M{C}^{(j)}_{\tilde{\V{\theta}},k}\big(\V{p}^{(p)}_k,\gamma_{k}^{(j,p)},1,\eta_{k}^{(j,p)}\big)$ and $\M{C}_{\tilde{\V{\theta}},n,0}^{(j,p)} \triangleq \M{C}^{(j)}_{\tilde{\V{\theta}},k}\big(\V{p}^{(p)}_k,\gamma_{k}^{(j,p)},0,\eta_{k}^{(j,p)}\big)$. 

\subsubsection{Learning Step (Unsupervised \ac{ai} Model Learning)}

The updated parametrization $\tilde{\V{\theta}}_{t+1}$ is determined by maximizing \eqref{eq:Q_particles}, i.e.,
\vspace*{-2mm}
\begin{align}
	\tilde{\V{\theta}}_{t+1} =\text{arg max}_{\tilde{\V{\theta}}}\iist \tilde{Q}(\tilde{\V{\theta}}| \tilde{\V{\theta}}_t).
	\label{eq:costfunction}\\[-6mm]\nn
\end{align}

\section{Data-driven Models}
\label{sec:data_model}

The data-driven components introduced in the measurement model in Section~\ref{subsec:meas_model} parameterize environment-induced multipath contributions via the mappings in \eqref{eq:nnpos} and \eqref{eq:nnrho}. These mappings are implemented using \acp{dnn} and jointly define a position-dependent generative model of the \ac{rf} propagation environment. The concept is shown in Fig~\ref{fig:factor-graph}.

\paragraph{\ac{nn} Inputs} For each \ac{bs}~$j$, the \acp{nn} are conditioned solely on the \ac{bs} position $\V{p}^{(j)} \in \mathbb{R}^d$, which serve as the input to the model. This parametrization allows the network to capture site-specific propagation characteristics and environment-dependent effects associated with each \ac{bs} location, while remaining independent of the instantaneous \ac{mt} position. If required, $\V{p}^{(j)}$ can be augmented with positional encodings to better represent fine-grained spatial variations, yielding an effective input dimension of $D_\text{in} = d + N_\text{enc} d$.

\paragraph{\ac{nn} Architecture and Outputs} Given the input $\V{p}^{(j)}$, the \acp{nn} predict the parameters of the environment-induced multipath model via the mappings $f^{\text{\scriptsize ai}}_{\V{p},\V{\theta}}(\V{p}^{(j)}) \in \mathbb{R}^{d+1 \times D}$ in \eqref{eq:nnpos} and $f^{\text{\scriptsize ai}}_{\rho,\V{\theta}}(\V{p}^{(j)}) \in \mathbb{R}^{1 \times D}$ in \eqref{eq:nnrho}. These outputs define the predicted map features $\V{P}^{\text{\scriptsize ai}\ist(j)}_{\V{\theta}}$, from which the corresponding delays, \acp{doa}, and power-related variances $\V{\lambda}^{\text{\scriptsize ai}\ist(j)}_{\V{\theta}}$ of the $D$ environment-induced \acp{mpc} are obtained and used to directly parameterize the covariance matrix in \eqref{eq:fullCov}. Specifically, $f^{\text{\scriptsize ai}}_{\V{p},\V{\theta}}(\V{p}^{(j)})$ is implemented by a three-layered \ac{mlp} with \texttt{ReLU} activation functions and hidden-layer dimensions $\{L_{\V{p},1},L_{\V{p},2},d \times 3D\}$. To guarantee strictly positive biases, $b^{\text{\scriptsize ai}\ist(j)}_{\V{\theta},n}$ is constrained via the absolute-value mapping $|\cdot|$. Likewise, $f^{\text{\scriptsize ai}}_{\rho,\V{\theta}}(\V{p}^{(j)})$ is implemented as a three-layer \ac{mlp} with hidden dimensions $\{L_{\V{p},1}, L_{\V{p},2}\}$ and output size $D$. Strictly positive real-valued variances are enforced by applying the absolute value $|\cdot|$ to the outputs of $f^{\text{\scriptsize ai}}_{\rho,\V{\theta}}(\V{p}^{(j)})$.

\paragraph{Learning Schemes} All \ac{nn} parameters $\V{\theta}$ and unknown array response parameters $\V{\chi}$ are learned in an unsupervised manner using the cost function in \eqref{eq:costfunction}. The \ac{nn} parameters are updated to $\V{\theta}_{t+1}$ while keeping $\V{\chi}=\V{\chi}_t$ fixed, with $\V{\theta}_t$ and $\V{\chi}_t$ used to evaluate the cost function. Conversely, the array response parameters are updated to $\V{\chi}_{t+1}$ while keeping $\V{\theta}=\V{\theta}_t$ fixed, again using the same cost function.

\subsubsection{\Ac{mt} Track Segmentation and Online Learning}

Learning can be performed using the full track $\V{z}_{1:K}$ ($K$ is the last time step of a track) or shorter track segments $\V{z}_{k-k_0:k}$. Accordingly, the parameters $\tilde{\V{\theta}}$ are obtained by maximizing the accumulated objective in \eqref{eq:Q_particles2} by summing $\tilde{Q}_k(\tilde{\V{\theta}}|\tilde{\V{\theta}}_t)$ over the corresponding time indices and using an \texttt{Adam}-optimizer to learn the parametrizations of the \acp{nn}.

\subsubsection{Approximate Representations}

To reduce the computational complexity of the learning step, we employ two approximation schemes:
(1) only a subset of $P_0 < P$ particles is retained after resampling, and
(2) the particle-based beliefs of the \ac{mt} state $\V{x}_k^{(p)}$ and the latent variables $\gamma_k^{(j,p)}$ and $\eta_k^{(j,p)}$ are replaced by their \ac{mmse} estimates given in \eqref{eq:MTmmseptcl}, \eqref{eq:losmmseptcl}, and \eqref{eq:noisemmseptcl}, respectively.
Furthermore, if ground-truth \ac{mt} positions $\V{x}_k^{\text{gt}}$ are available, partially supervised learning can be performed by conditioning the cost function on $\V{x}_k^{\text{gt}}$ instead of $\hat{\V{x}}_{k,\mathrm{MMSE}}$.

\subsubsection{Learning of Response Parameters $\V{\chi}$ (optional)}

Since $\V{\chi}$ enters the likelihood function in \eqref{eq:likelihood} through the response vector $\V{h}_{\V{\chi}}(\cdot)$, the gradient of the cost function in \eqref{eq:Q_particles2} with respect to $\V{\chi}$ is used to learn these parameters via \texttt{Adam}-optimizer, while keeping $\V{\theta}=\V{\theta}_t$ fixed. This enables data-driven calibration of the response vector $\V{h}_{\V{\chi}}(\cdot)$ in \eqref{eq:arrayres} directly from raw measurements.

\section{Results}
\begin{figure}[t]
	\centering
	\setlength{\abovecaptionskip}{0pt}
	\setlength{\belowcaptionskip}{0pt}
	\setlength{\figurewidth}{0.28\textwidth}
	\setlength{\figureheight}{0.28\textwidth}
	\tikzsetnextfilename{scenario}
	\scalebox{1}{\includegraphics{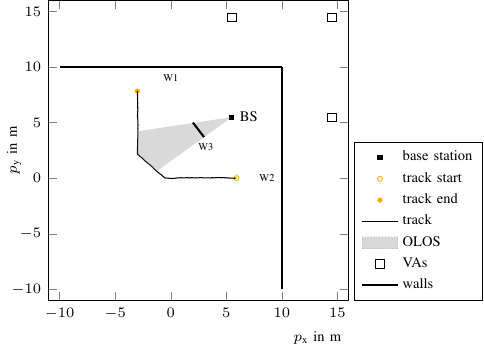}}
	\caption{Graphical representation of the investigated synthetic experiment: Fig. (a) shows the simulated trajectory, the (single) known \ac{bs} position, unknown walls (and respective \acfp{va}) and indices the \ac{olos} situation. 
	}\label{fig:scenario}
	\vspace{-6mm}
\end{figure}

We compare (i) the proposed \ac{ai}-enhanced direct \ac{slam} method with (ii) a direct \ac{los}-only \ac{spa} (the \ac{ai}-induced covariance term in \eqref{eq:fullCov} is deactivated) and (iii) a genie-aided direct \ac{spa} (true map features are used in covariance matrix in \eqref{eq:fullCov}, cf.~\cite{LeiMeyMeiWitHla:GNSS2016}). The simple multipath propagation environment and chosen \ac{ai} model are used solely to demonstrate that the unsupervised approach can learn the generative signal model (map features and amplitude statistics) in a controlled scenario; the framework and \ac{ai} components can be flexibly scaled without altering the learning principle.

\subsubsection{Simulation Setup and Scenario} \label{sec:simulation_setup}

We evaluate the proposed algorithm using synthetic radio measurements that were generated using the measurement model in \eqref{eq:radiosigmod}. We use the simple scenario presented in Fig.~\ref{fig:scenario}, where the \ac{mt} receives radio signals from only a single \ac{bs}, while moving along an unsteady track with two distinct direction changes. It is observed at $190$ discrete time steps $k$ at a constant observation rate of $100\,\mathrm{ms}$.
The ground truth \ac{mpc} distances are calculated based on the simple floor plan of Fig.~\ref{fig:scenario} (W1 and W2) using the image source or \ac{va} model \cite{Ped:TAP2019,LeiVenTeaMey:TSP2023} considering single bounce and double bounce reflections.  
The amplitudes of the LOS component as well as the \acp{mpc} are assumed to follow free-space path loss according to their individual propagation paths, and determined by a SNR of $42~\mathrm{dB}$ at a distance of $1~\mathrm{m}$. The amplitudes of \acp{mpc} are additionally attenuated by 3 dB per reflection. 
The \ac{bs} is obstructed by an obstacle (W3), which leads to an \ac{olos} situation in the center of the track.
We choose the transmitted signal to be of root-raised-cosine shape with a roll-off factor of $0.6$ and a $3$-dB bandwidth of $500\,\mathrm{MHz}$ at a center frequency of $f_\mathrm{c} = 7\,\mathrm{GHz}$. The received baseband signal is critically sampled, i.e., $T_\text{s} = 1.25\,\mathrm{ns}$, with a total number of $N_\text{s} =81$ samples, amounting to a maximum distance $d_\text{max} = 30\,\mathrm{m}$. We use a uniform rectangular array with $M_\text{a}=4$ spaced at $\lambda/2$ with $\lambda=c/f_\mathrm{c}$.
The state transition \ac{pdf} $f(\bm{x}_n|\bm{x}_{n-1})$ of the \ac{mt} state $\V{x}_n$ is described by a linear, constant-velocity and stochastic-acceleration model\cite[p.~273]{BarShalomBook:Book2001}, given as $\V{x}_n = \bm{A}\, \V{x}_{n- 1} + \bm{B}\, \V{w}_{n}$ 
with the acceleration process $\V{w}_n$ being i.i.d. across $n$, zero mean, and Gaussian with covariance matrix ${\sigma_{\text{a}}^2}\, \bm{I}_2$, % = 
the acceleration standard deviation ${\sigma_{\text{a}}}$, and $\bm{A} \in \mathbb{R}^{\text{4x4}}$ and $\bm{B} \in \mathbb{R}^{{4\times 2}}$ being defined according to \cite[p.~273]{BarShalomBook:Book2001}. The acceleration standard deviation is set to $\sigma_a=2~\mathrm{m/s^2}$ assuming that it is estimated by an IMU sensor. The particles for the initial \ac{mt} position and velocity at $k=0$ are drawn i.i.d.\ from Gaussian distributions, whose mean for each realization is itself sampled from a Gaussian centered at the true \ac{mt} state, with initialization standard deviations of $0.5~\text{m}$ and $0.1~\text{m/s}$. The state-transition parameters for the amplitude and noise states are set to $c_{\gamma}=100$ and $c_{\eta}=100$. The numbers of particles are $P=5000$ and $P_\text{a}=250$, respectively. The appearance and visibility probabilities are $p_\text{a}=0.01$ and $p_\text{v}=0.95$, and the appearance prior $f_{\mathrm{a}}(\gamma_{k}^{(j)})$ is uniform on $[0,2]$. The initial distributions for noise variance $f(\eta^{(j)}_{0})$ and \ac{los} amplitude variance $f(\gamma^{(j)}_{0})$ for each \ac{bs} are uniformly distributed on $[0,\ist 5\cdot10^{-4}]$ and $[0,\ist 2]$, respectively. The initial LOS existences probability is $f(r^{(j)}_{0}) = 0.5$ for each \ac{bs}.

For implementation, we use Python with \texttt{PyTorch} and the \texttt{Adam} optimizer with default parameters ($\beta_1=0.9$, $\beta_2=0.999$, $\epsilon=10^{-8}$). Separate learning rates are used for the two networks, i.e., $10^{-4}$ for the amplitude \ac{nn} and $5\cdot 10^{-3}$ for the map-feature \ac{nn}. Unsupervised updates are performed every $k_0=30$ time steps (track segment) using $P_0=30$ particles drawn from the respective marginal posterior \acp{pdf}, with $300$ back-propagation iterations per segment. The number of \ac{ai} components $D =30$. To prevent an initially concentrated map and encourage exploration during early online updates, the neural map features $\V{p}^{\text{\scriptsize ai}\ist(j)}_{\V{\theta},n}$ are initialized such that their decoded means are uniformly distributed over a predefined region ($[-35\,\text{m}, 35\,\text{m}] \times [-35\,\text{m}, 35\,\text{m}]$).\footnote{	
	%Specifically, after standard initialization of the hidden layers of $f^{\text{\scriptsize ai}}_{\V{p},\V{\theta}}(\cdot)$, 
	The final linear layer is calibrated by evaluating the \ac{bs} inputs $\V{p}^{(j)}$ and solving a least-squares fit from the last hidden features to uniformly sampled features $\V{P}^{\text{\scriptsize ai}\ist(j)}_{\V{\theta}}$ in \eqref{eq:nnpos}, yielding map features that span the area of interest while preserving the network architecture.	
	}

\begin{figure}[t]
	\centering
	\setlength{\belowcaptionskip}{0pt}	
	\tikzsetnextfilename{results}
	\vspace{-2mm}
	\includegraphics{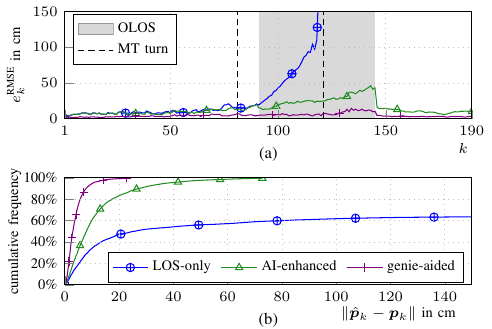}
	\vspace{-2mm}
	\caption{Performance in terms of the \ac{rmse} of the estimated \ac{mt} position over time $n$ (a) and as the cumulative frequency of the magnitude error of the estimated \ac{mt} position (b). The gray area represents the area of \ac{olos} between  \ac{bs} and \ac{mt} according to Fig.~\ref{fig:scenario}.
	}\label{fig:results}
	\vspace{-6mm}
\end{figure}

\subsubsection{Numerical Experiment} \label{sec:experiment}
Fig.~\ref{fig:results} presents the numerical results in terms of the \ac{rmse} of the estimated \ac{mt} position over time,
$
e_{k}^{\text{RMSE}} = \mathbb{E}_{\bm{p}_k}[\norm{\hat{\bm{p}}_k - \bm{p}_k}^{2}]^{1/2},
$
and the cumulative distribution of the position error magnitude, averaged over $50$ simulation runs.
Without any prior floorplan information, the proposed method learns the feature map and amplitude statistics from the geometric information obtained in the initial phase of the track. Hence, it can maintain accurate localization in the \ac{olos} situation despite the single \ac{bs} setup, with the error staying below $75$ cm in all realizations, where the direct \ac{los}-only algorithm diverges. Its performance approaches that of the genie-aided direct \ac{spa}, which serves as a lower bound in the considered scenario. Since the proposed particle-based implementation can be executed in parallel over particles $P$ and anchors $J$, we used a GPU-based implementation. The mean runtime of the AI-enhanced algorithm was $30~\mathrm{ms}$ per time $k$ (even for $D =30$) for the \textit{inference step} and $250~\mathrm{ms}$ for the \textit{unsupervised track-segment learning step}.

\section{Conclusion}
We presented an \ac{ai}-enhanced direct \ac{slam} method that combines Bayesian inference on raw \ac{rf} signals with unsupervised environment learning. The proposed method combines model-based components (capturing the \ac{los} contribution) with learned \ac{ai} components (capturing the environment-induced \acp{mpc}) in order to retain a physically consistent signal model that explicitly models the sampled \ac{rf} signal spectrum and array response and that learns the environment in terms of \ac{mpc} statistics and propagation geometry. 
The proposed \ac{elbo} maximization enables principled unsupervised training of the generative environment model without requiring labeled positions or map features. Numerical results in a simple multipath scenario demonstrate that the learned model effectively captures environment geometry and amplitude statistics and, thus, supports accurate \ac{mt} localization, especially under \ac{olos} conditions.  
Potential directions for future research include introducing partially modeled map features and hierarchical feature models \cite{LiaLeiMey:TSP2025, XuhongTWC2026}, incorporating uncertainty-aware environment representations in order to improve generalization and validating performance on real-world measurements.

\renewcommand{\baselinestretch}{0.95}\small\normalsize
\bibliographystyle{ieeetr}
\bibliography{IEEEabrv,StringDefinitions,Papers,Books}

\end{document}